\documentclass{PoS}
\pdfoutput=1

\usepackage[tbtags]{amsmath}

\usepackage{wrapfig}
\usepackage{graphicx}

\newcommand{\bc}{\begin{center}}
\newcommand{\ec}{\end{center}}
\newcommand{\ms}{\mskip 1.5mu}

\def\t0{t\text{{\ttfamily =}}0}
\def\ix0{\xi\text{{\ttfamily =}}0}
\newcommand{\Dlr}{{D^{\hspace{-0.8em}%
      \raisebox{0.8ex}{$\scriptstyle\leftrightarrow$}}}{}}

\title{Progress in hadron structure physics on the lattice\thanks{preprint \textrm{TUM-T39-07-15}}}

\ShortTitle{Progress in hadron structure physics on the lattice}

\author{\speaker{Philipp H\"agler}\\
        Institut f\"ur Theoretische Physik T39,
   Physik-Department der TU M\"unchen, James-Franck-Stra\ss{}e, 85747
   Garching, Germany\\
        E-mail: \email{phaegler@ph.tum.de}}

\abstract{ This is a review of progress in hadron structure physics from lattice QCD.
Recent results on the structure of the nucleon and the pion in terms of (transition) form factors, moments
of distribution amplitudes and (generalized) parton distribution functions are presented. These observables allow us to investigate a number of fundamental physics questions related to e.g. the distribution of charge and momentum in hadrons, the spin structure of the nucleon and the pion, and correlations between spin, orbital angular momentum and coordinate degrees of freedom. Chiral extrapolations of selected lattice results are presented and compared to results from experiment and phenomenology. We conclude that lattice simulations already today strongly contribute to our understanding of the structure of hadrons.}

\FullConference{The XXV International Symposium on Lattice Field Theory\\
		 July 30 - August 4 2007\\
		 Regensburg, Germany}

\begin{document}

\section{Introduction}
The past year has seen impressive 
efforts with respect to dynamical lattice QCD calculations of many hadron structure observables
like form factors and moments of (generalized) parton distribution functions, which are by now 
consistently carried out at pion masses as low as $300$ MeV.  
New methods and techniques have been developed and
successfully tested in e.g. calculations of the form factor and gluonic structure of the pion, 
exploratory studies of disconnected and strange quark contributions
to nucleon structure observables, as well as a first investigation of the neutron electric 
polarizability with dynamical quarks. 
Other new and interesting developments include a study of recent lattice results on
nucleon to $\Delta$ axial-vector transition form factors
in the small scale expansion of chiral perturbation theory,
an exploratory study of moments of vector meson distribution amplitudes, 
which represent an essential part in the understanding of rare $B$ decays,
and a calculation of moments of tensor generalized parton distributions of the pion, 
giving rise to a surprisingly non-trivial pion spin structure.

Many of these observables can be defined using bi-local quark operators on the light-cone,
\begin{eqnarray}
\label{Op}
  O_\Gamma(x)=
\int \frac{d \lambda}{4 \pi} e^{i \lambda x}
  \overline{q} \big(-\frac{\lambda}{2}n\big)\!
  \Gamma \mathcal{U}
  q\big(\frac{\lambda}{2} n\big)  ,
\end{eqnarray}
where the variable $x$ represents the longitudinal quark momentum fraction, $n$ is a light cone vector,
$\Gamma=\gamma^\mu,\gamma^\mu\gamma_5,\sigma^{\mu\nu}\ldots$, and the Wilson-line $\mathcal{U}$ 
ensures gauge invariance.
Lattice calculations give access to hadron matrix elements of $x$-moments of $O_\Gamma(x)$.
The moments are obtained by taking the integral $\int_{-1}^{1} dx x^{n-1}$, leading to a tower of local operators
\begin{equation}
 \label{localOps}
  \mathcal{O}_{\gamma}^{\mu\mu_1\cdots\mu_{n-1}}
= \mathcal{S}\; \overline{q} \, \gamma^{\mu}\ms
      i\Dlr^{\mu_1} \cdots\ms i\Dlr^{\mu_{n-1}}\ms q \,,
\end{equation}
for the vector case $\Gamma=\gamma^\mu$, where $\mathcal{S}$ denotes symmetrization in the indices 
$\mu,\mu_1,\ldots$
and subtraction of traces. Similar expressions can be obtained for the axial-vector,
$\Gamma=\gamma^\mu\gamma_5$, and tensor,
$\Gamma=\sigma^{\mu\nu}$, operators.
In QCD, operators as given in Eq.~(\ref{localOps}) have to be renormalized and therefore 
lead in general to scale and scheme dependent quantities. Typical examples of corresponding lattice operators
for $n=2$ are discussed below in section \ref{sec:nuclmomfrac}.
Non-forward matrix elements $\langle P'|\ldots|P\rangle$ of Eq.~(\ref{Op}) can be parametrized 
in terms of generalized parton distributions (GPDs), e.g. $H(x,\xi,t)$ and $E(x,\xi,t)$ for the 
nucleon in the vector case,
\begin{eqnarray}
   \langle P'|O_{\gamma}^{\mu}(x)|P\rangle
   &=&\overline{U}(P')\bigg(\gamma^{\mu}H(x,\xi,t) +\frac{i\sigma^{\mu\rho}\Delta_\rho}{2m_N}E(x,\xi,t) \bigg)U(P)\,,
   \label{MEs1}
\end{eqnarray}
where $t=\Delta^2=(P'-P)^2$ is the momentum transfer squared, and $\xi$ corresponds to the
longitudinal momentum transferred to the hadron (for reviews on GPDs, see \cite{Diehl:2003ny,Belitsky:2005qn}).
Note that $H(x,\xi,t)$ reduces in the forward limit, $\Delta\rightarrow 0$, to the usual unpolarized parton
distribution function, $H(x,0,0)=q(x)$.
Corresponding nucleon matrix elements of the local operators in Eq.~(\ref{localOps})
for $n=1,2$ are given by
%
\begin{eqnarray}
   \langle P'|\mathcal{O}_{\gamma}^{\mu}|P\rangle
   &=&\overline{U}(P')\bigg(\gamma^{\mu}F_1(t) +\frac{i\sigma^{\mu\rho}\Delta_\rho}{2m_N}F_2(t) \bigg)U(P) \,,
   \nonumber\\
   \langle P'|\mathcal{O}_{\gamma}^{\mu\mu_1}|P\rangle
    &=&\mathcal{S}\;\overline{U}(P')\bigg(\gamma^{\mu}\overline{P}^{\mu_1}A_{20}(t)   
     +\frac{i\sigma^{\mu\rho}\Delta_\rho}{2m_N}\overline{P}^{\mu_1}B_{20}(t) 
     +\frac{\Delta^{\mu}\Delta^{\mu_1}}{m_N}C_{20}(t)\bigg)U(P) \,,
   \label{MEs}
\end{eqnarray}
where $\overline{P}=(P'+P)/2$, and $F_{1,2}$ are the Dirac and Pauli form factors. 
Since $\mathcal{O}_{\gamma}^{\mu\mu_1}$ corresponds to the quark contribution to the QCD energy momentum
tensor, the generalized form factors (GFFs) 
$A_{20}(t)$ and  $B_{20}(t)$  at $t=0$ can be 
directly related to the quark momentum fraction,
$\langle x\rangle_q=A_{20}(0)$, and the total quark 
angular momentum, $J_q=(A_{20}(0)+B_{20}(0))/2$.
The lattice calculation of matrix elements as in Eq.~(\ref{MEs}) for the nucleon from
two- and three-point functions, and the subsequent extraction of
moments of parton distributions and (generalized) form factors follows standard methods described in e.g. 
\cite{Dolgov:2002zm,Hagler:2003jd,Gockeler:2003jfa} and references therein.
Some new techniques and strategies used in the calculation of hadron structure observables on 
the lattice will be discussed in the sections below.
\section{Form factors}
\subsection{Axial-vector coupling constant}
\begin{figure}[t]
\bc
\includegraphics[scale=.7,clip=true,angle=0]{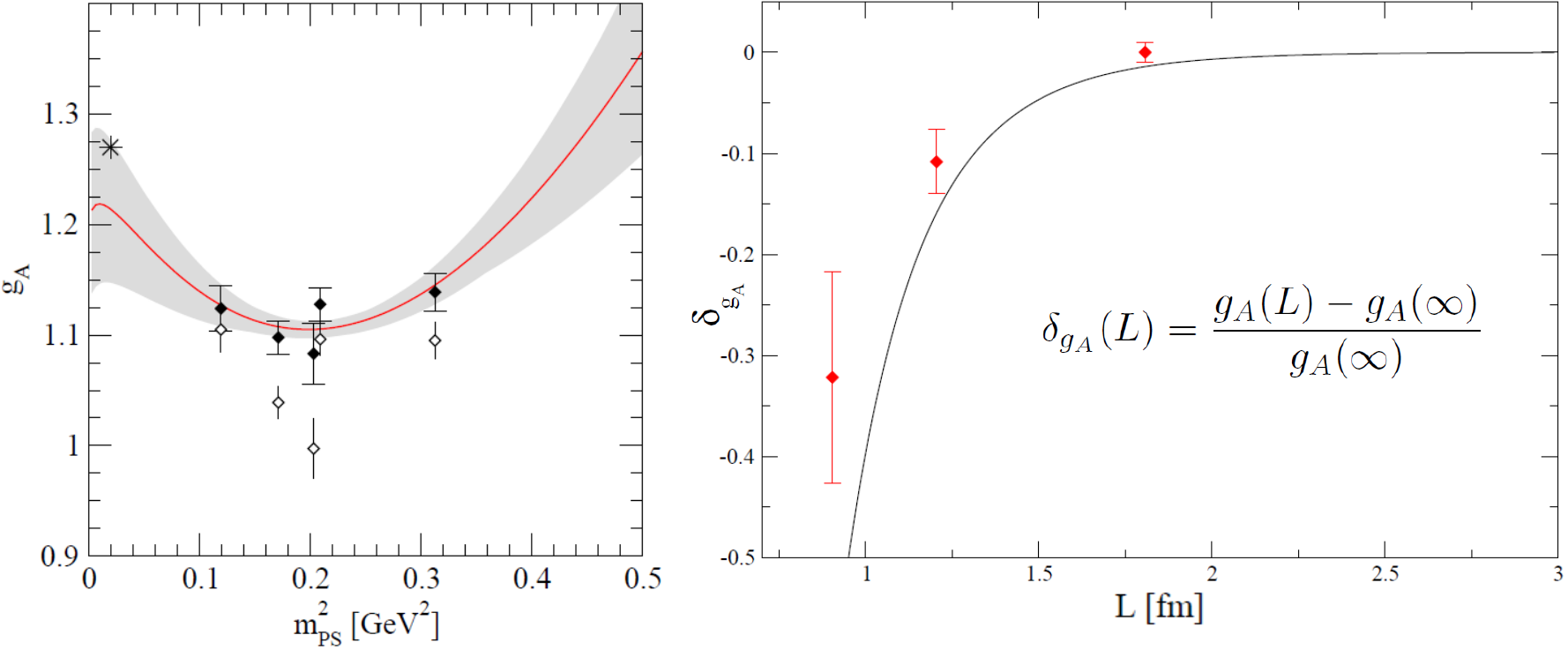}
  \caption{Results for $g_A$ from QCDSF/UKQCD \cite{Pleiter:PoSLat2007}. Symbols are explained in the text.}\label{gAQCDSF}
  \ec 
  \vspace*{-0.5cm}
\end{figure}
The axial-vector coupling constant ("axial charge") of the nucleon, $g_A=1.2695(29)$, 
is experimentally very well known from neutron beta decay. It corresponds to the forward limit of
the isovector axial-vector form factor, $G_A(Q^2\rightarrow 0)=g_A$. 
In a lattice calculation, using isospin symmetry, $g_A$ can be extracted from 
the proton forward matrix element of the $u-d$ axial-vector current,
\begin{equation}
\langle P,S | \bar{u} \gamma_\mu \gamma_5 u - \bar{d} \gamma_\mu \gamma_5 d | P,S  \rangle
= \bar{U} (P,S) \gamma_\mu \gamma_5  U(P,S)\,g_A \,.
\label{Eq:gA}
\end{equation}
Equation (\ref{Eq:gA}) shows that $g_A$ is directly related to the isovector quark spin fraction, $g_A=\Delta\Sigma_{u-d}$,
and that disconnected diagrams, which cancel out for the flavor combination $u-d$, do not contribute. 
The chiral limit value of the axial-vector coupling, $g^0_A$, is a fundamental low energy constant
of the chiral effective field theory of QCD (ChEFT).
Based on the Adler-Weisberger sum rule, one might anticipate that the $\Delta$ resonance,
in addition to pion and nucleon degrees of freedom, plays an important role in the low energy description
of $g_A$ in the framework of chiral perturbation theory (ChPT). It is also known that
the axial-vector coupling is particularly sensitive to finite volume effects in, e.g., a lattice simulation.
The application of ChPT including the $\Delta$ resonance in a finite volume to the description and
extrapolation of lattice data on $g_A$ therefore seems to be very promising.  
Figure \ref{gAQCDSF} shows new preliminary results from QCDSF/UKQCD for $g_A$, 
based on $N_f=2$ improved Wilson fermions and Wilson glue, for pion masses as 
low as $\simeq340$ MeV \cite{Pleiter:PoSLat2007}. The operator has been non-perturbatively renormalized
using the Rome-Southampton method. The scattered open symbols on the left already indicate 
that finite size effects (FSEs) may be large. 
Indeed, a fit based on ChPT including explicitly the $\Delta$ resonance in a finite volume
\cite{Hemmert:2003cb,Beane:2004rf,Khan:2006de,Procura:2006gq}  
reveals a significant dependence on the lattice extent, $L$,
see the RHS of Fig.~\ref{gAQCDSF} for a fixed pion mass of $\approx 600$ MeV, 
and allows for a simultaneous description of the $m_\pi$ and $L$ dependence
of the lattice data.
\begin{wrapfigure}{r}{0.45\textwidth}
  \vspace*{-.1cm}
  \begin{center}
    \includegraphics[width=0.43\textwidth]{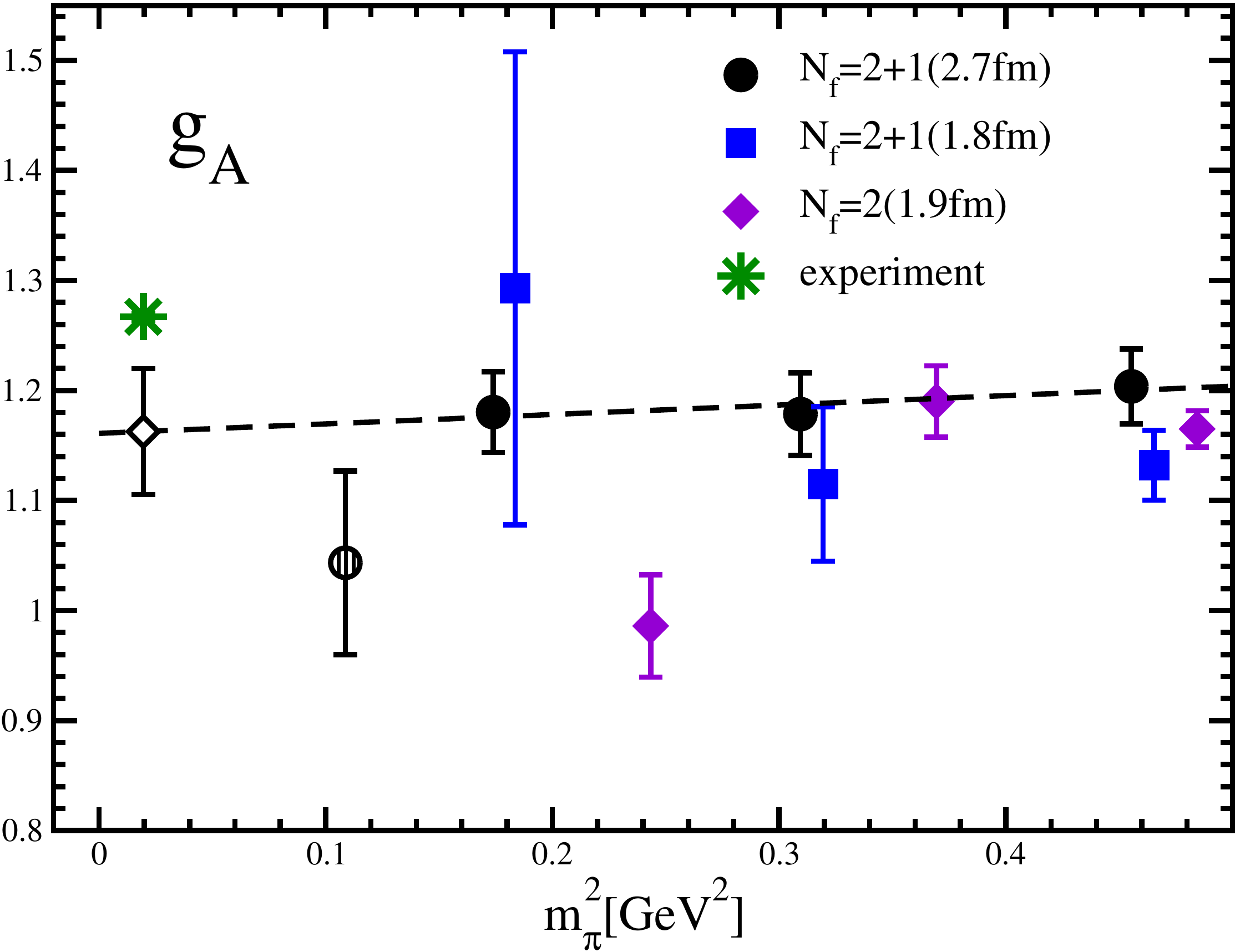}
  \end{center}
  \vspace*{-0.5cm}
  \caption{Results for $g_A$ from RBC-UKQCD\cite{Yamazaki:PoSLat2007}.}
  \label{gARBC}
 \end{wrapfigure}
\noindent The lattice results and the 
corresponding chiral fit, projected to the infinite volume, 
are represented by the filled symbols and the shaded error band,
respectively, on the LHS of Fig.~\ref{gAQCDSF}. The curvature of the chiral extrapolation is just strong enough
as to provide an overlap with the experimental value at the physical point, indicated by the star.
Preliminary results on $g_A$ by RBC-UKQCD \cite{Yamazaki:PoSLat2007} based on $N_f=2+1$ domain wall
fermions with $L_s=16$ and volumes of $1.8$ and $2.7$ fm are shown in Fig.~\ref{gARBC}. 
It is gratifying to see that dynamical chiral fermion calculations at pion masses as low as $\simeq330$ MeV
lead to results for $g_A$ with statistical errors around the 5\% level. However, the significantly lower
result at the lowest pion mass, represented by the shaded circle, may indicate that FSEs are substantial.
For comparison with results from LHPC at similar pion masses see \cite{Edwards:2005ym}.

\subsection{Nucleon charge radii and anomalous magnetic moments}
\begin{figure}[t]
   \begin{minipage}{0.48\textwidth}
      \centering
          \includegraphics[width=.9\textwidth,clip=true,angle=0]{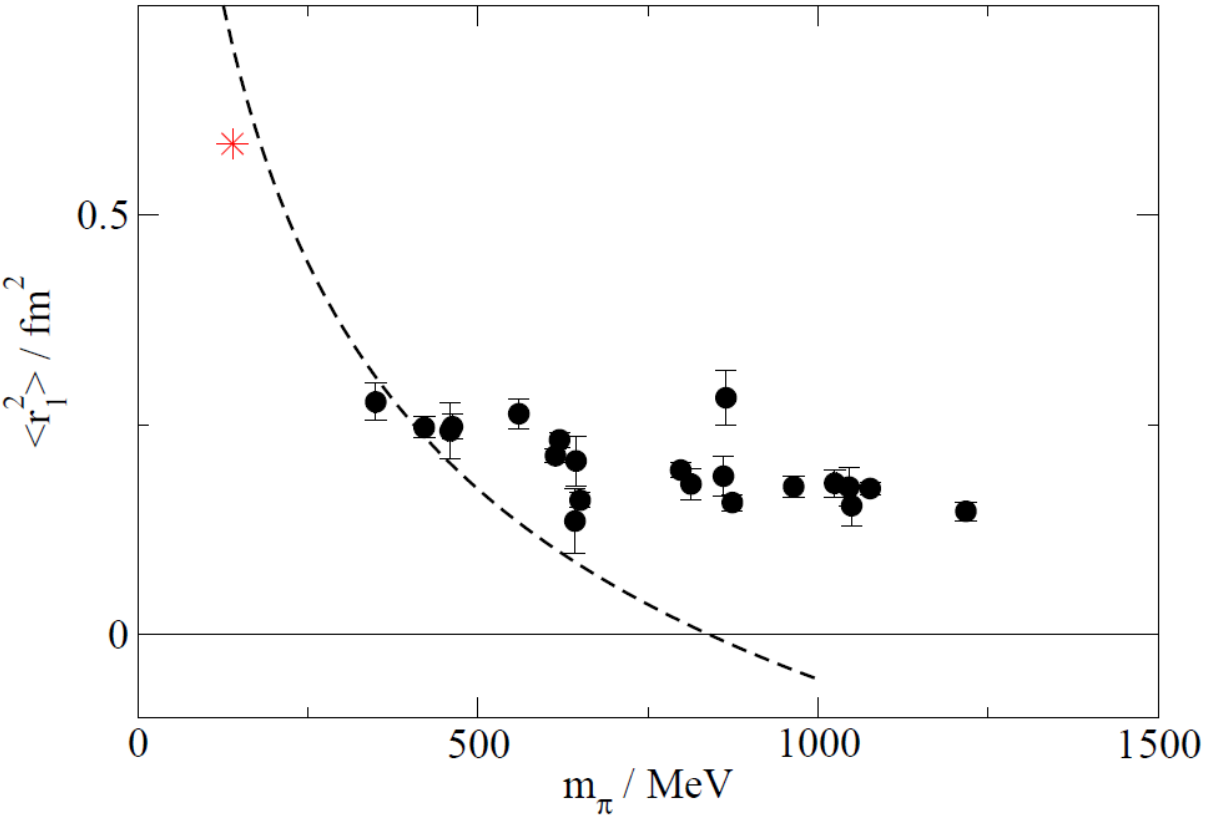}
  \caption{Isovector mean square radius $\langle r^2\rangle_1$ from QCDSF/UKQCD \cite{Schroers:PoSLat2007}.\newline}\label{r2QCDSF}
     \end{minipage}
     \hspace{0.5cm}
    \begin{minipage}{0.48\textwidth}
      \centering
          \includegraphics[width=0.8\textwidth,clip=true,angle=0]{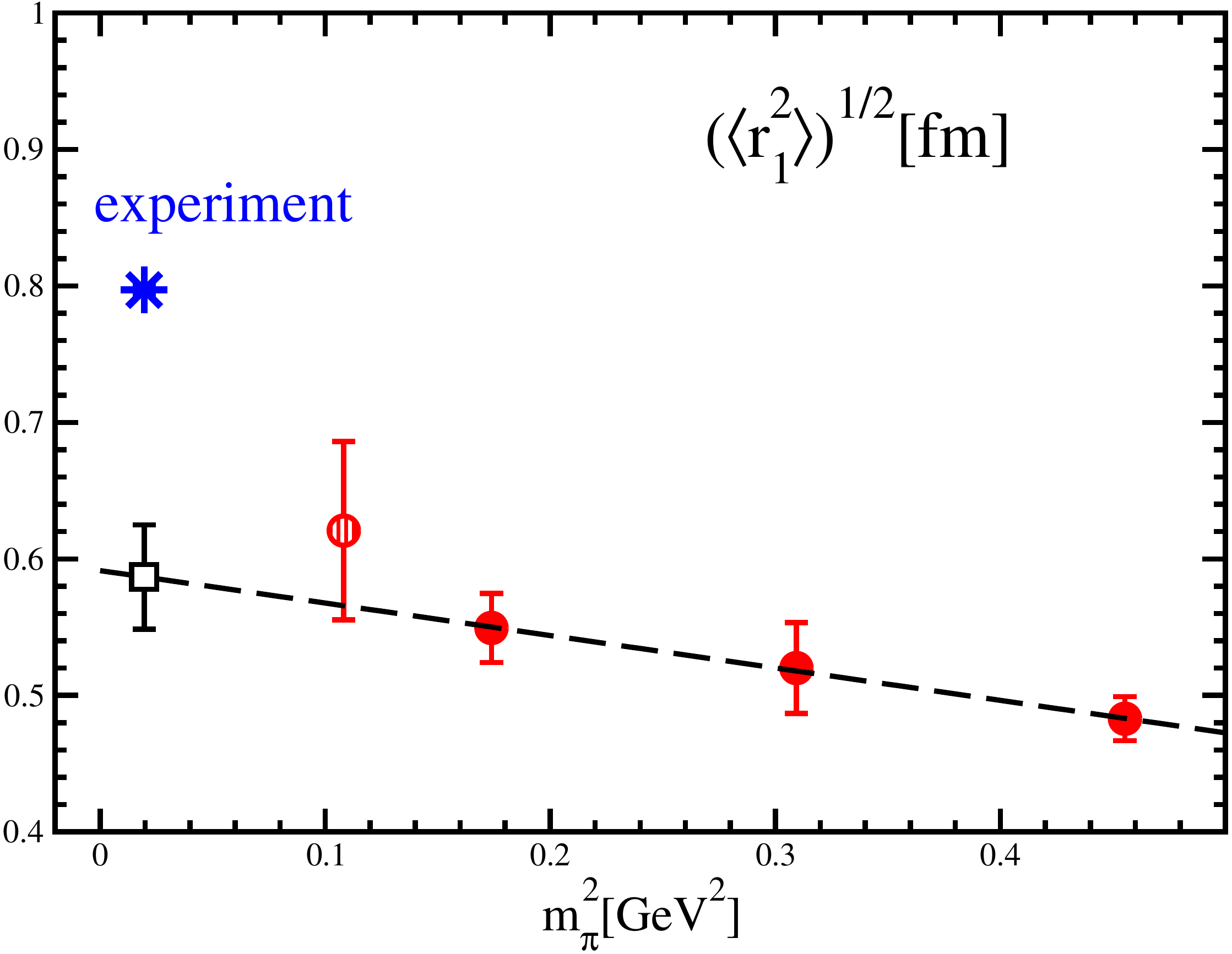}
  \caption{Isovector root mean square radius $\langle r^2\rangle^{1/2}_1$ from RBC-UKQCD \cite{Yamazaki:PoSLat2007}.\newline}\label{r2RBC}
     \end{minipage}
 \end{figure}

Despite interesting developments on the theoretical as well as the experimental side in this field, publications on nucleon form factors (FFs) $F_{1,2}(Q^2)$ 
in unquenched lattice QCD are surprisingly scarce (for recent reviews, see \cite{Perdrisat:2006hj,Arrington:2006zm}).
This includes such fundamental observables as mean square charge radii, $\langle r^2\rangle$, 
and the anomalous magnetic moment, $\kappa$, 
\begin{eqnarray}
\langle r^2\rangle_i&=&\frac{-6}{F_i(Q^2=0)}\left[\frac{d}{dQ^2}F_i(Q^2)\right]_{Q^2=0}\,,\label{Eq:r2}\\
\kappa&=&F_2(Q^2=0)=\mu-F_1(Q^2=0)\,,
\label{Eq:kappa}
\end{eqnarray}
where $i=1,2$ and $\mu=G_M(Q^2=0)$ is the nucleon magnetic moment.
Preliminary results on $\langle r^2\rangle^{u-d}_1$ as a function of the pion mass 
by QCDSF/UKQCD \cite{Schroers:PoSLat2007} are shown in Fig.~\ref{r2QCDSF}.
The mean square isovector charge radius has been obtained from a new parametrization \cite{Schroers:PoSLat2007} 
of the $Q^2$-dependence of the lattice results for the Dirac FF $F_1(Q^2)$.
Even at the lowest pion mass of $\simeq340$ MeV, the lattice data points are almost a factor
of two below the experimental result, which is represented by the star. Chiral perturbation theory in the 
form of the small scale expansion (SSE) to $\mathcal{O}(\epsilon^3)$ in the infinite volume \cite{Gockeler:2003ay} 
predicts a rather strong slope in $m_\pi$, which barely connects the leftmost lattice datapoints with the 
experimental number, as indicated by the dashed line.
It will be highly interesting to see if the lattice results follow the chiral extrapolation curve at
pion masses below $300$ MeV. 

\begin{figure}[t]
\vspace*{-0.5cm}
\bc
 \includegraphics[scale=0.65,clip=true,angle=0]{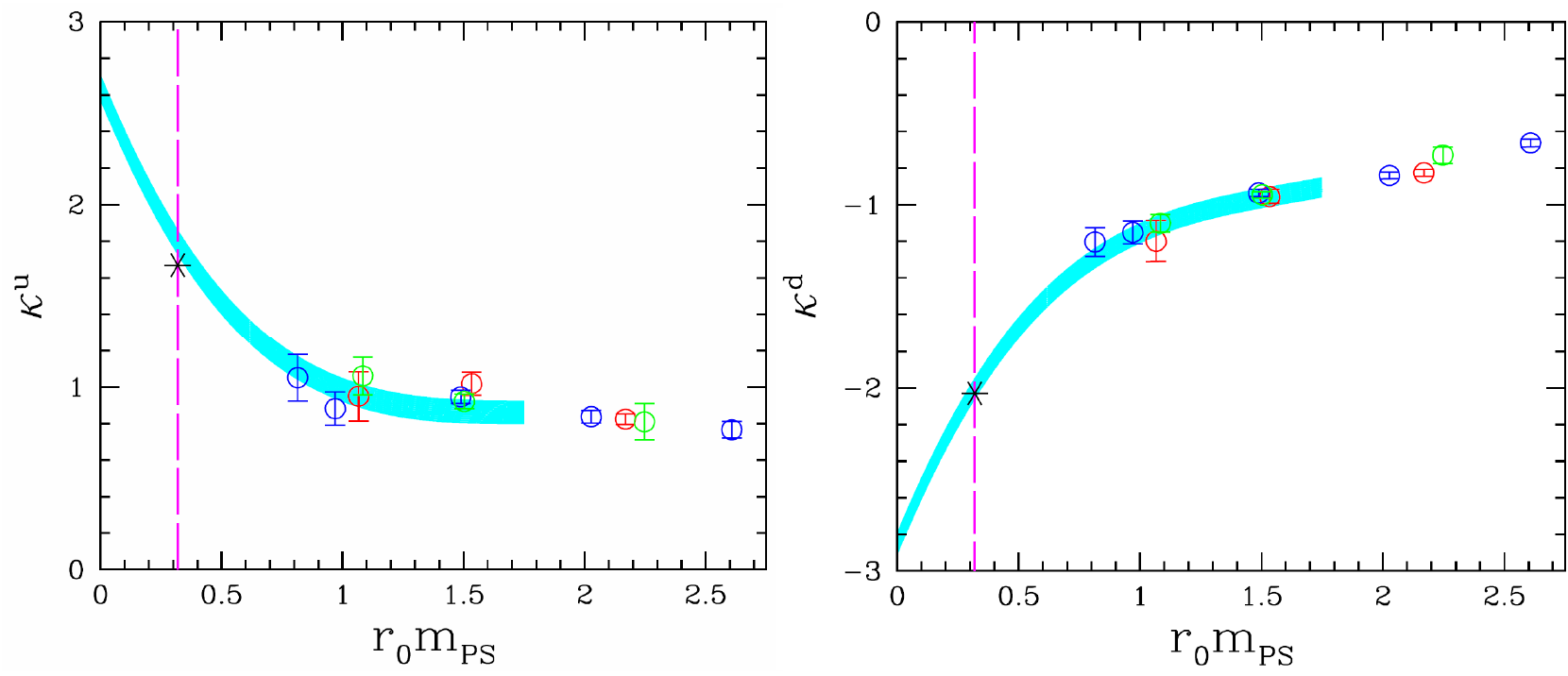}
    \caption{Quark contributions to the nucleon anomalous magnetic moment $\kappa$ from QCDSF/UKQCD \cite{Schroers:PoSLat2007}.}
  \label{kappaQCDSF}
\ec 
 \vspace*{-0.5cm}
\end{figure}

Similar results have been obtained by RBC-UKQCD \cite{Yamazaki:PoSLat2007}, Fig.~\ref{r2RBC},
where the charge radius has been extracted from a dipole fit to the 
lattice results, i.e. $\langle r^2\rangle^{u-d}_1=12/m_D^2$, with the dipole mass $m_D$. 
Figure~\ref{kappaQCDSF} shows preliminary results from QCDSF/UKQCD \cite{Schroers:PoSLat2007}
on the up- and down-quark contributions
to the nucleon anomalous magnetic moment. The results of 3-parameter chiral fits based on the 
small scale expansion (SSE) \cite{Gockeler:2003ay}
to the lattice datapoints are represented by the shaded errors bands, 
showing good agreement with the experimental values, indicated by the stars, at the physical point. 
Nevertheless, these results should be taken with due caution since contributions from disconnected diagrams have not been included.

Further results on vector and axial-vector nucleon form factors, including the axial-vector charge radius 
and the pseudoscalar coupling constant, have been presented by QCDSF/UKQCD \cite{Schroers:PoSLat2007}, RBC-UKQCD
\cite{Yamazaki:PoSLat2007} and the Athens-Cyprus-MIT collaboration \cite{Alexandrou:2007xj,Tsapalis:PoSLat2007}.

\subsection{Pion form factor}
\begin{figure}[t]
     \begin{minipage}{0.4\textwidth}
      \centering
          \includegraphics[scale=.52,clip=true,angle=0]{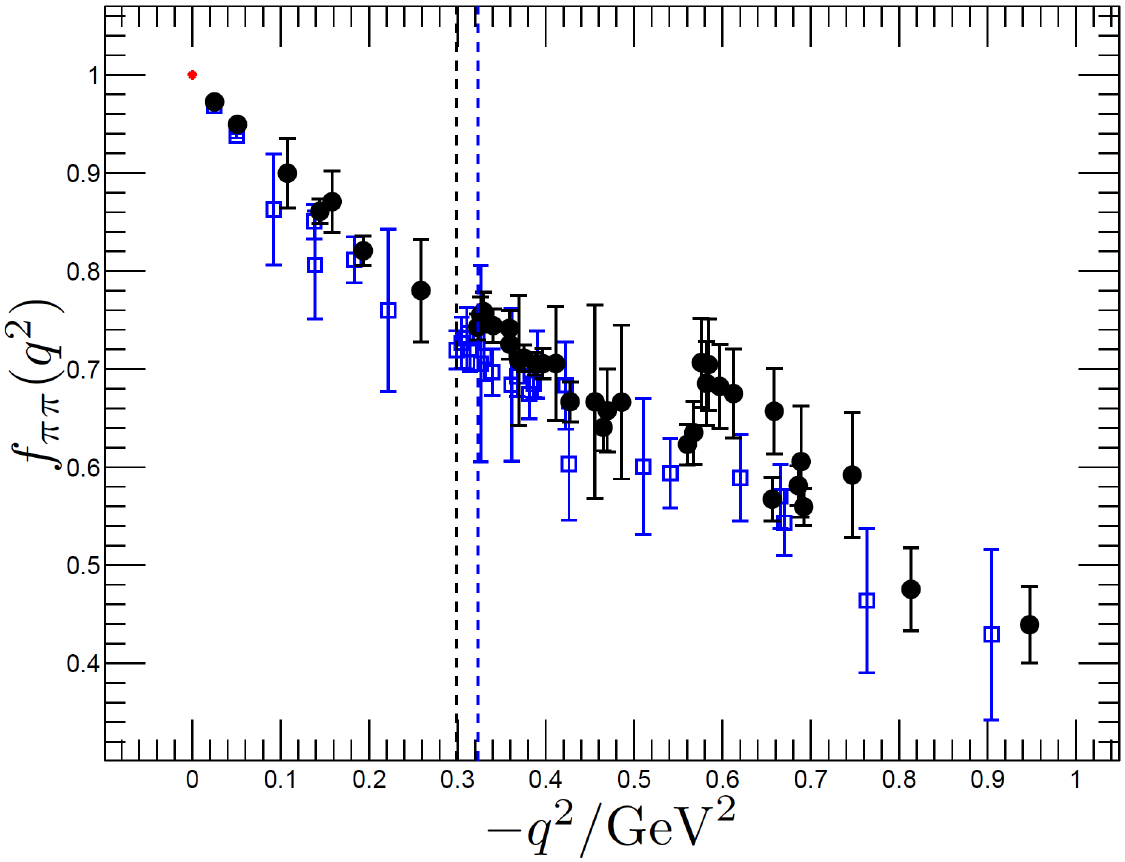}
  \caption{Pion FF using pTBCs from \cite{Boyle:2007wg}.\newline}\label{pionFFSouth}
     \end{minipage}
     \hspace{0.cm}
    \begin{minipage}{0.6\textwidth}
      \centering
       \vspace*{-0.2cm}
        \includegraphics[width=0.9\textwidth,angle=0]{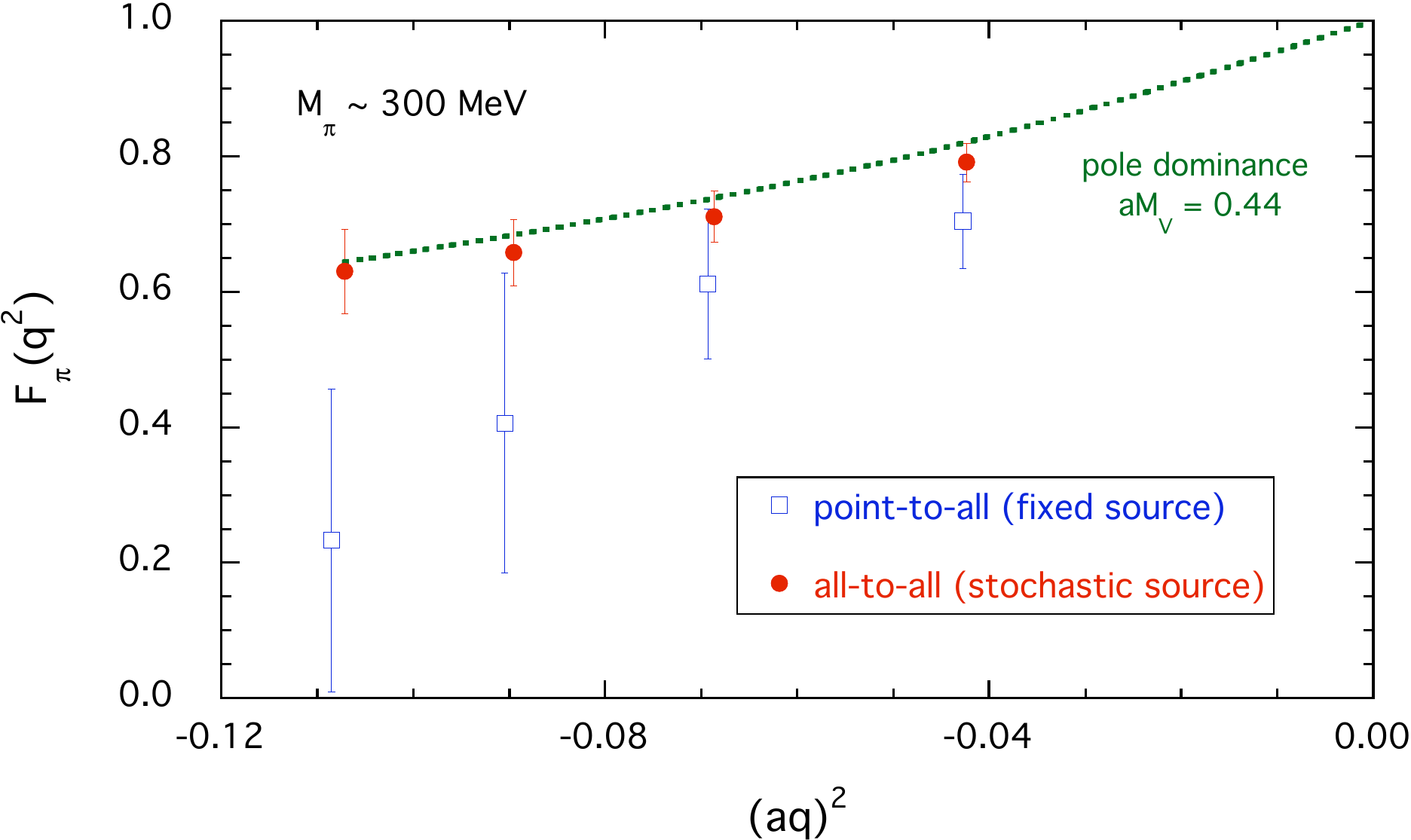}
        \vspace*{-0.2cm}
  \caption{Pion FF from ETMC \cite{Simula:PoSLat2007}.\newline}\label{pionFFETMC}
     \end{minipage}
     \vspace*{-0.5cm}
    \end{figure}

Substantial progress based on new methods and techniques can be seen 
in lattice calculations of, e.g., the pion form factor $F_\pi(Q^2)$. 
The analysis of many important hadron structure observables requires small but non-zero
values for the (squared) momentum transfer $q^2=(p_f-p_i)^2=-Q^2$. Typical examples are charge radii (\ref{Eq:r2}),
anomalous magnetic moments (\ref{Eq:kappa}) and angular momentum contributions to the nucleon spin 
(see section \ref{SecSpin}),
where in the latter two cases the corresponding (generalized) FFs cannot be obtained at $Q^2=0$ because
their contribution to hadron matrix elements vanishes in the limit $Q^2\rightarrow 0$. This poses a
problem for current lattice calculations, since due to limited spatial lattice extents, $L=16,\ldots,48$, 
the lowest available non-zero lattice momenta are $p=2\pi/(aL)\approx 300$ MeV. A way around this
in a dynamical lattice calculation is to use partially twisted boundary conditions (pTBCs) 
\cite{Sachrajda:2004mi,Bedaque:2004ax}, leading
to a modified expression for the momentum transfer,  
$q_{TBC}^2\!=\!(E_{f,TBC}-E_{i,TBC})^2\!-\!(\vec q+(\vec\theta_f-\vec\theta_i)/(aL))^2$ with 
$E_{i(f),TBC}=((\vec p_{i(f)}+\vec\theta_{i(f)}/(aL))^2+m^2)^{1/2}$,  where  
the twisting angles $(\theta_{i(f)})_{j=1,2,3}$ can be tuned continuously. 
An exploratory study of this technique, based
on $N_f=2+1$ DW fermions, has been published recently \cite{Boyle:2007wg}, 
and the results for the pion FF are shown in Fig.~\ref{pionFFSouth}.
The vertical dashed lines indicate the lowest non-zero $q^2$ which can be obtained in the conventional
calculation, and all lattice datapoints to the left of these lines (except at $q^2\!=\!0$)
could only be obtained using \mbox{pTBCs}. The smallness of the error bars for non-zero twisting angle
but vanishing initial and final hadron three-momenta (two leftmost filled squares) is impressive, in particular
in comparison with the error bars of datapoints obtained for non-zero hadron momenta, which are
a factor of 5 to 10 larger due to larger statistical noise. 
Similarly noteworthy progress has been reported by ETMC in the calculation of $F_\pi(Q^2)$ 
based on twisted mass Wilson fermions \cite{Simula:PoSLat2007} at pion masses as low as $300$ MeV. 
All-to-all propagators have been evaluated using stochastic sources and 
utilized for the calculation of pion three- and two-point-functions,  
leading to a substantial improvement of the precision compared to the standard approach employing point-to-all
propagators, see Fig.~\ref{pionFFETMC}.

\begin{wrapfigure}{r}{0.45\textwidth}
  \begin{center}
   \vspace*{-0.3cm}
      \includegraphics[width=0.45\textwidth]{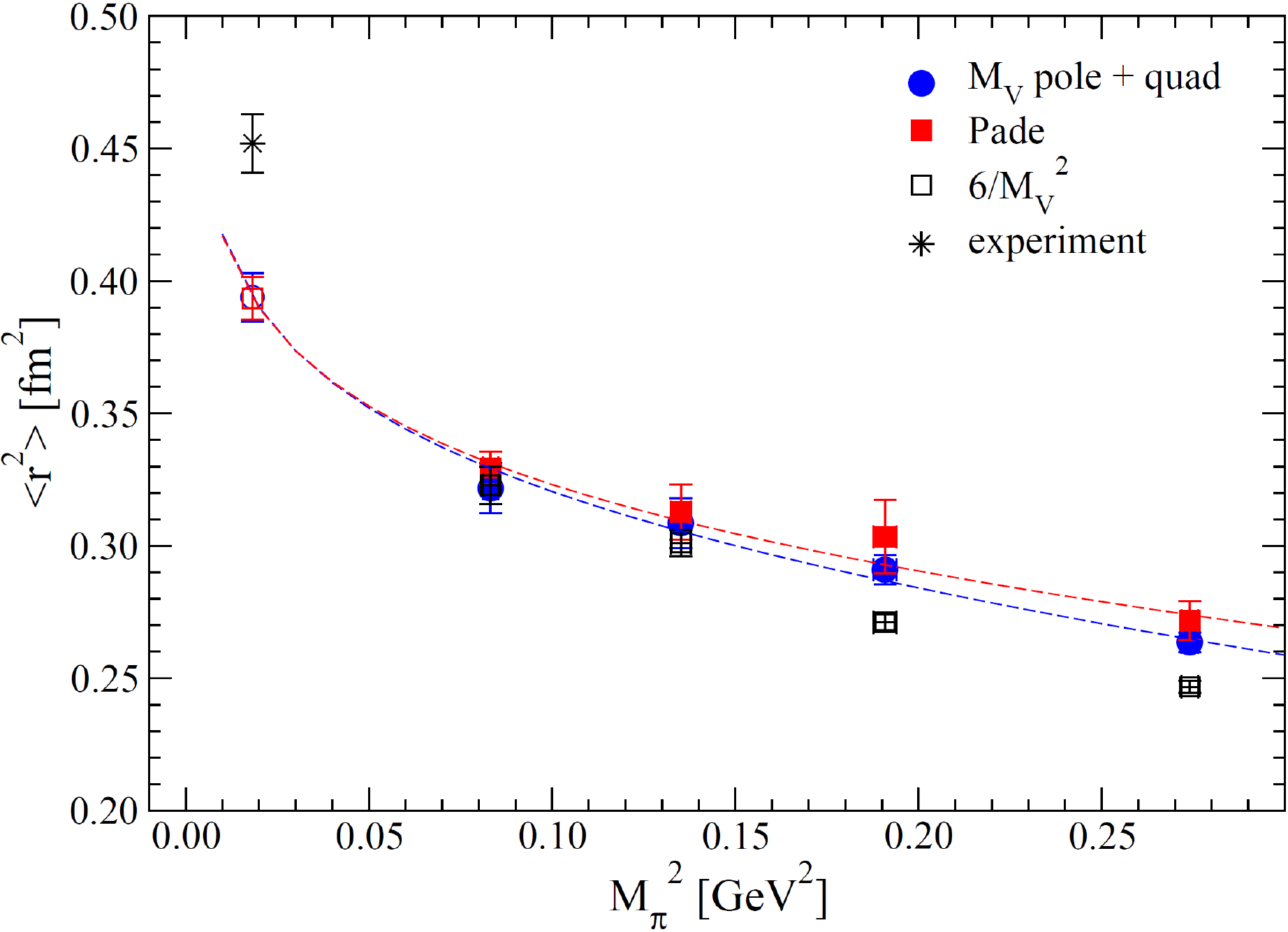}
  \end{center}
   \vspace*{-0.8cm}
  \caption{Pion mean square charge radius from JLQCD \cite{Kaneko:PoSLat2007}.}
   \vspace*{-0.3cm}
  \label{pionFFJLQCD}
\end{wrapfigure}
A calculation of the pion FF at pion masses down to $288$ MeV, using $N_f=2$ overlap fermions with a Wilson kernel
and the Iwasaki gauge action in a fixed topological sector, has been presented by JLQCD \cite{Kaneko:PoSLat2007}. 
The calculation of the pion correlation functions is based on all-to-all propagators, 
which have been evaluated following the strategy proposed in \cite{Foley:2005ac}.
Preliminary results for the pion charge radius, obtained from a variety of parametrizations 
of the $Q^2$-dependence of $F_\pi(Q^2)$, versus the pion mass are shown in Fig.~\ref{pionFFJLQCD}. It is remarkable
that a statistical precision at the few percent level has been achieved with chiral fermions in a dynamical
calculation at such low pion masses.
Further studies of systematic effects related to e.g. the
finite lattice volume, setting of the scale and fixed topology may help to understand why
the absolute values for $\langle r^2\rangle_\pi$ in Fig.~\ref{pionFFJLQCD} are rather low
compared to experiment, and also lower than the results obtained recently in an
extensive study of $F_\pi(Q^2)$ based on $N_f=2$ improved Wilson fermions \cite{Brommel:2006ww}.

\subsection{Vector meson quadrupole moments}
Matrix elements of the vector current for spin one hadrons,
e.g. $\langle \rho(P',S') | \bar{q} \gamma_\mu q | \rho(P,S)\rangle$ for the $\rho$, 
can be parametrized by the three form factors $G_E(Q^2)$, $G_M(Q^2)$ and $G_Q(Q^2)$. Their
forward values at $Q^2=0$ are directly related to the electric charge $q$, the magnetic
moment $\mu$, and the quadrupole moment $Q$, respectively, e.g. $G_Q(Q^2=0)=m_\rho^2 Q_\rho$.
All these observables obviously yield important information about hadron structure, and
quadrupole moments in particular can exhibit possible spatial deformations of hadrons.
Recently, the vector meson form factors have been studied in quenched lattice QCD using
fat-link irrelevant clover (FLIC) fermions at pion masses as low as $\approx290$ MeV \cite{Hedditch:2007ex}.
As in the case of magnetic moments, quadrupole moments cannot be directly extracted from 
the lattice correlators, since their contribution to the matrix elements 
vanishes in the limit $Q^2\rightarrow 0$. Assuming that the $Q^2$-dependence of the
electric and quadrupole form factors is similar at low $Q^2$, $G_Q(Q^2)/G_E(Q^2)\approx \text{const.}$, 
one obtains an approximation for the quadrupole moment of, e.g., the $\rho$ from
$Q_\rho\approx G_Q(Q^2)/(G_E(Q^2)m_\rho^2)$ for small $Q^2$. Based on results 
for $G_E(Q^2)$ and $G_Q(Q^2)$ at $Q^2\simeq0.22$ GeV$^2$, a clearly non-zero small negative value,
$Q_\rho\approx-0.007$ fm$^2$, has been obtained for the $\rho$ quadrupole moment
\cite{Hedditch:2007ex}. Together with results for the charge radius
$\langle r^2\rangle_\rho$ of the $\rho$, this leads to a ratio of $|Q_\rho|/\langle r^2\rangle_\rho\approx1/30-1/50$,
indicating that the $\rho$ is not completely spherically symmetric but slightly oblate.

\subsection{Nucleon to $\Delta$ axial-vector transition form factors}
\begin{figure}[t]
     \begin{minipage}{0.5\textwidth}
      \centering
          \includegraphics[scale=.6,clip=true,angle=0]{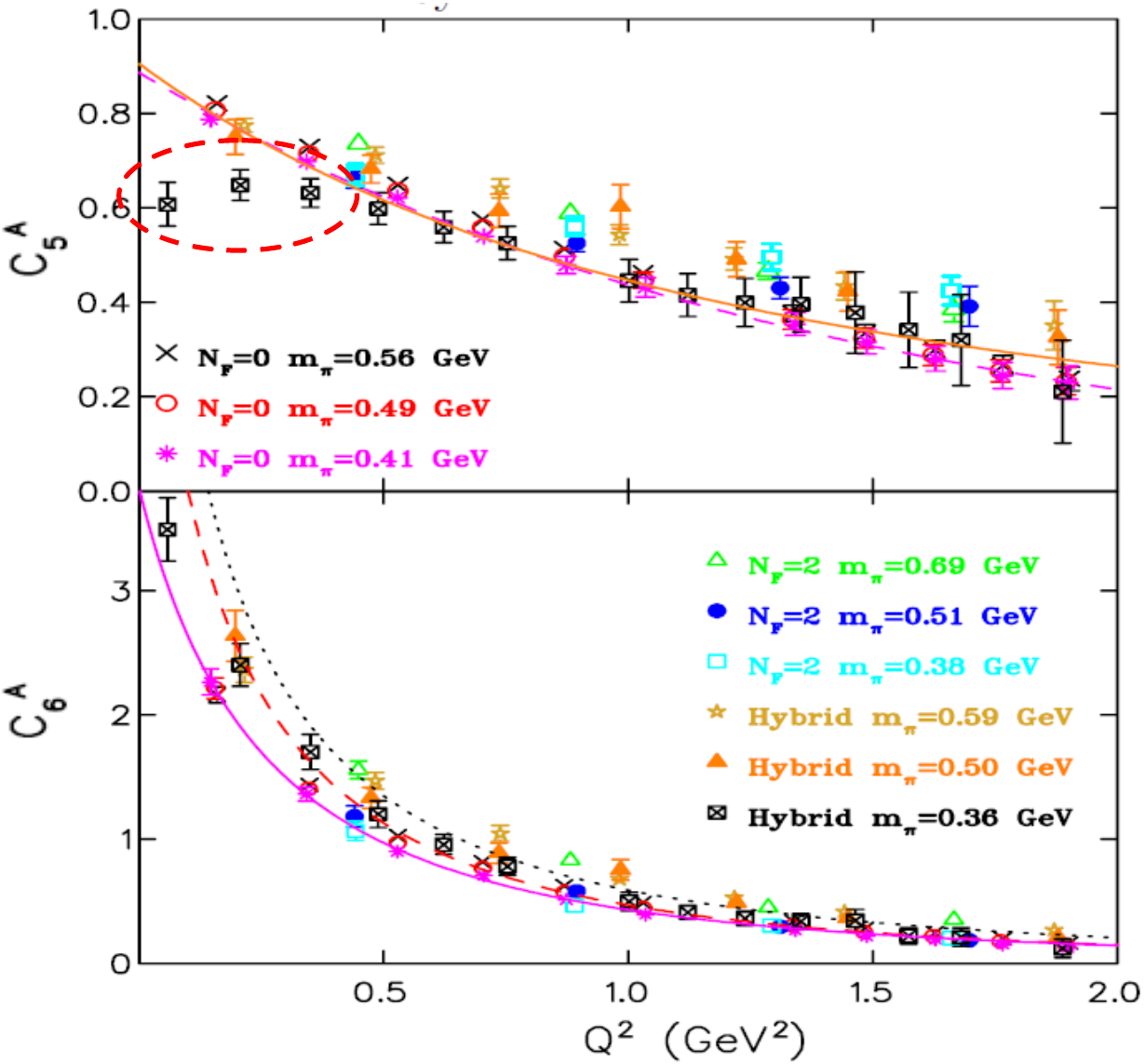}
  \caption{The transition FFs $C^A_{5,6}$ calculated in \cite{Alexandrou:2007xj}.\newline}\label{CA5Cyprus}
     \end{minipage}
     \hspace{0.2cm}
     \begin{minipage}{0.5\textwidth}
      \centering
          \includegraphics[width=0.8\textwidth,clip=true,angle=0]{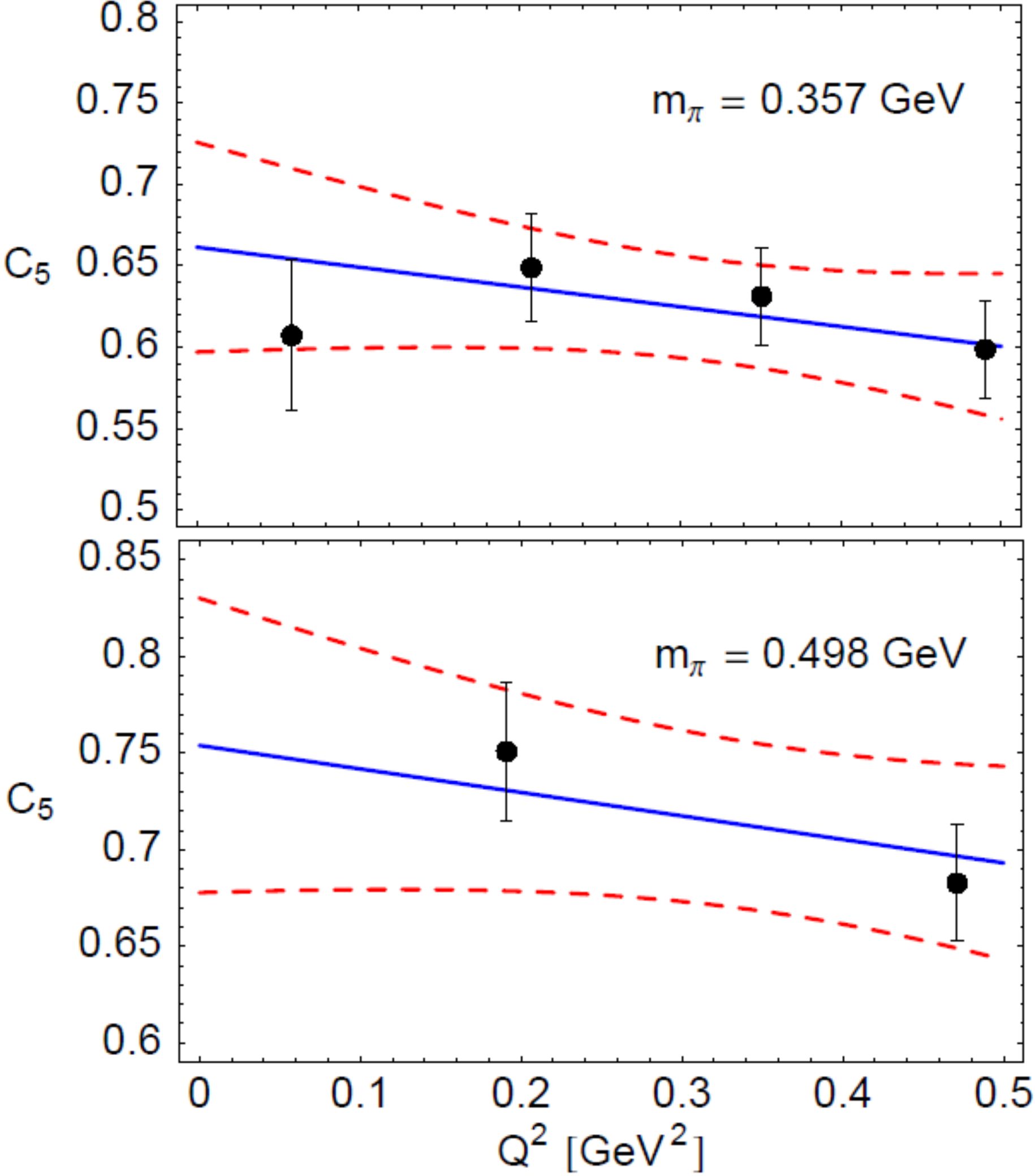}
  \caption{$C^A_5$ in ChPT (SSE) by Procura \cite{Procura:xyz2007}.}
  \label{CA5Procura}
     \end{minipage}
   \end{figure}
In contrast to the vector mesons discussed in the previous section, the nucleon as a spin $1/2$ particle
has no (static) quadrupole moment which could be utilized to measure possible deviations from spherical symmetry.
Instead, spin $1/2$ to spin $3/2$, in particular nucleon to $\Delta$, vector transitions can 
and have been used to study possible non-zero quadrupole amplitudes, see, e.g., \cite{Bernstein:2007zz} 
and references therein.
Recently, the corresponding \emph{axial-vector} nucleon to $\Delta$ transition form factors 
have been investigated in detail in lattice simulations based on 
Wilson fermions in the quenched and unquenched theory
and in a hybrid approach of $N_f=2+1$ DW valence quarks on top of AsqTad 
staggered sea quarks (MILC configurations) \cite{Alexandrou:2007xj,Tsapalis:PoSLat2007}.
The results for the two dominant transition form factors, $C^A_5(Q^2)$ and $C^A_6(Q^2)$, 
are shown in Fig.~\ref{CA5Cyprus}. Apart from the overall very good statistical precision and
agreement between the different lattice actions, three lattice data points at low $Q^2$
and the lowest pion mass $\sim360$ MeV (crossed squares) stand out in the sense that they do not
follow the general trend of the data. In addition to studying possible systematic uncertainties
of the calculation, it is certainly very interesting to see if this downwards bending at low $Q^2$ and $m_\pi^2$
can be understood within ChPT. This has been looked at recently
by Procura \cite{Procura:xyz2007} in the framework of the small scale expansion (SSE)
to leading one-loop accuracy $\mathcal{O}(p^3)$. The result for $C^A_5$ is, in short hand form, given by
$C^{A,\text{SSE}}_5(Q^2,m_\pi)=C^{A,0}_5 + C^{A,m_\pi}_5 m_\pi^2 + C^{A,Q}_5 Q^2 + C^{A,loop}_5(m_\pi,c_A,g_A,g_1,f_\pi,\Delta)$, showing to this order a linear dependence on $Q^2$, while the
non-analytic dependence on $m_\pi$ is hidden in $C^{A,loop}_5$. The low energy constants 
in $C^{A,loop}_5$ may be fixed to known values from the literature and by a fit to lattice data for the axial-vector coupling \cite{Khan:2006de}. 
The result of a three parameter fit (with parameter $C^{A,0}_5$ plus the two counter terms $\propto Q^2$ and
$\propto m_\pi^2$) to six lattice data points of Fig.~\ref{CA5Cyprus} is presented in Fig.~\ref{CA5Procura}.
It is encouraging to see that, within the error bands indicated by the dashed lines, the SSE
calculation is compatible with the results from the hybrid lattice calculation.
\section{Polarizabilities in a dynamical calculation}
Nucleon electric and magnetic polarizabilities, $\alpha_E$ and $\beta_M$, 
parametrize the forward Compton scattering amplitude at $\mathcal{O}(\omega^2)$
in a low energy expansion in the photon energy $\omega$. 
They describe the resistance of the internal degrees of freedom of the nucleon against 
external forces created by the electromagnetic field of the scattered photon and 
therefore encode significant information on nucleon structure.
The polarizabilities are related to a shift of the hadron mass
through the effective Hamiltonian $H_{eff}=-(\alpha_E E^2+\beta_M B^2)/2$.

\begin{wrapfigure}{r}{0.5\textwidth}
\vspace*{-1.cm}
  \begin{center}
      \includegraphics[width=0.48\textwidth]{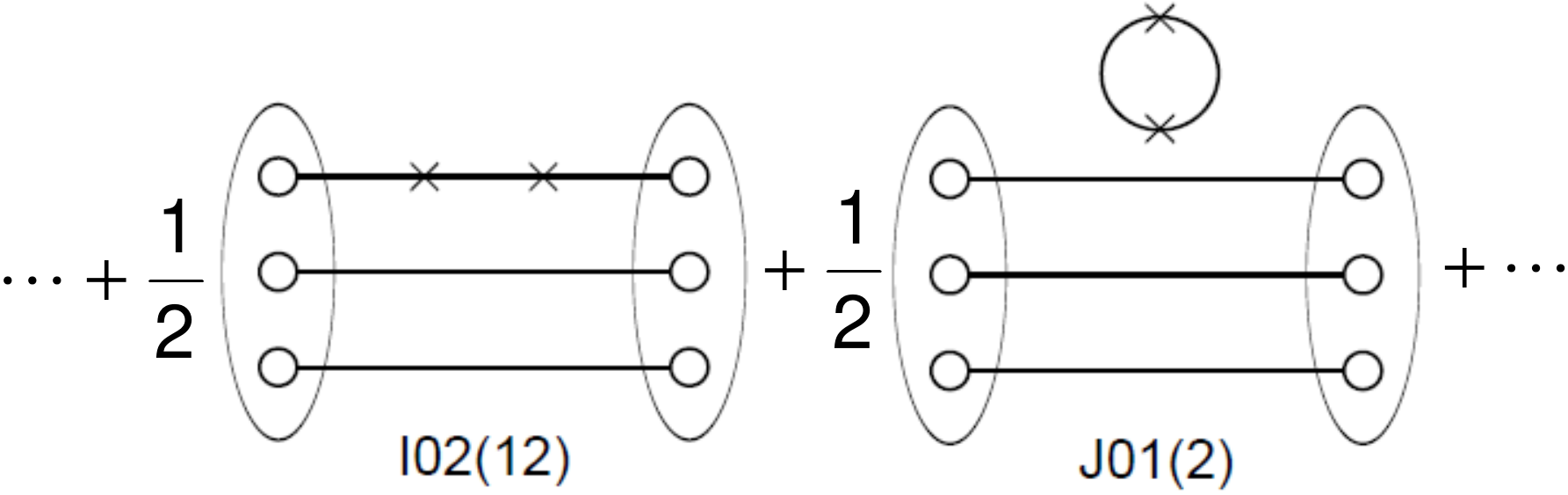}
  \end{center}
  \vspace*{-0.3cm}
 \caption{\label{Engel1} Examples for connected (left) and disconnected (right) four-point function contributions
to the neutron electric polarizability \cite{Engelhardt:2007ub}.}
\vspace*{-0.2cm}
\end{wrapfigure}
\noindent
In a recent study, Engelhardt \cite{Engelhardt:2007ub,Engelhardt:PoSLat2007} 
explored the feasibility of a new method to compute the electric polarizability 
of the neutron in a dynamical lattice calculation. The inclusion of
external fields in the calculation of gauge configurations renders such dynamical 
calculations at first sight prohibitively expensive.
\begin{figure}[t]
     \begin{minipage}{0.48\textwidth}
      \centering
          \includegraphics[width=0.8\textwidth,clip=true,angle=0]{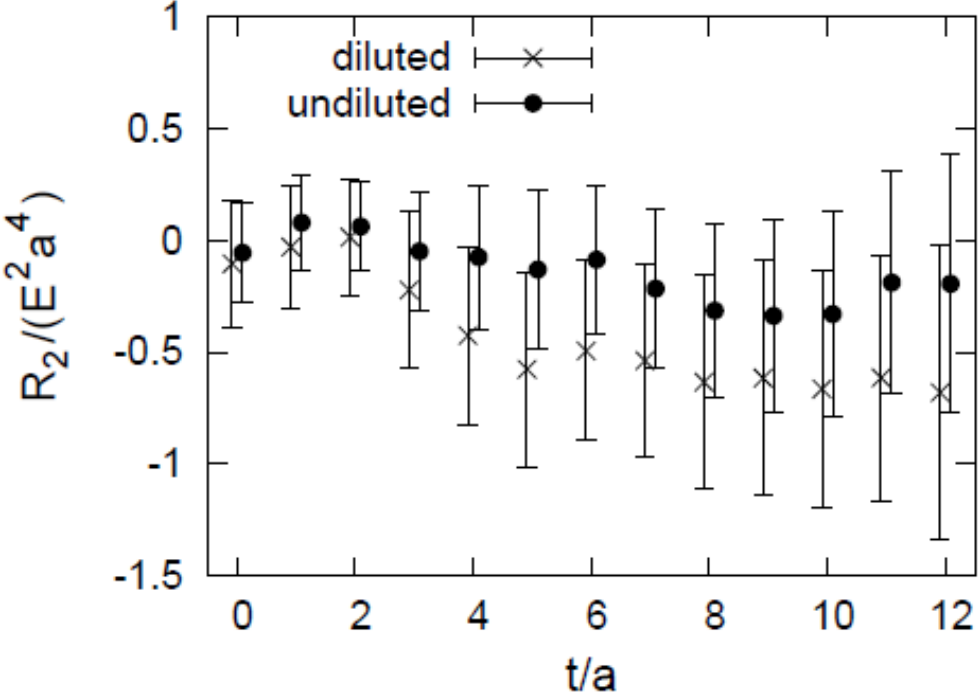}
  \caption{Contribution to the ratio $R_2$ from diagram J01 (see Fig.11) \cite{Engelhardt:2007ub}.}
  \label{Engel2}
     \end{minipage}
     \hspace{0.5cm}
     \begin{minipage}{0.48\textwidth}
      \centering
          \includegraphics[width=0.8\textwidth,clip=true,angle=0]{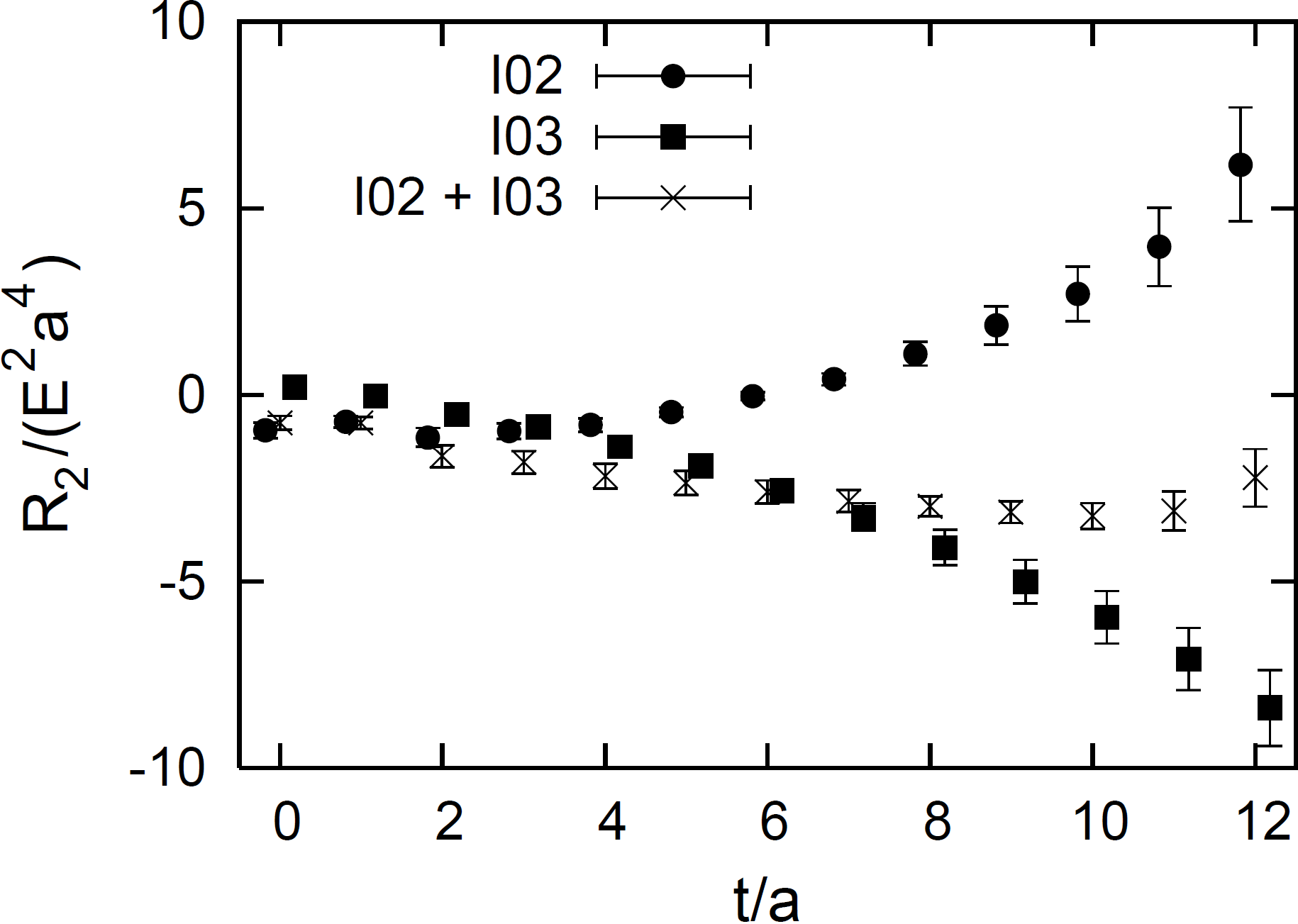}
  \caption{Contribution to the ratio $R_2$ from diagram I02 (see Fig.11) \cite{Engelhardt:2007ub}.}
  \label{Engel3}
     \end{minipage}
   \end{figure}
A way around this is to expand the neutron two-point function in the external (electric) field $E$
to the desired order, $\mathcal{O}(E^2)$, resulting in specific four-point functions. Examples
for connected and disconnected contributions are given in Fig.~\ref{Engel1}, 
where a cross represents an insertion linear in the external field. 
The mass shift, and thereby the electric polarizability, can then be extracted from the $t$-slope
of the ratio of the expanded two-point function to the two-point function for vanishing
external fields, $R_2(t)=G_{(E^2)}(t)/G_{(E=0)}(t)$, where $t$ is the source-sink separation
in Euclidean time direction. 
The numerical calculation has been based on the hybrid approach described 
in \cite{Edwards:2005ym,Hagler:2007xi}, i.e. 
domain wall valence fermions and a MILC gauge ensemble. 
Examples for ratios are shown in Fig.~\ref{Engel2} and  Fig.~\ref{Engel3} for a pion mass of $\simeq 760$ MeV.
The slope of $R_2$ for the disconnected contributions, which have been evaluated
using stochastic sources, is compatible with zero in Fig.~\ref{Engel2}, while 
non-zero slopes can be observed in Fig.~\ref{Engel3} for individual connected contributions.
In contrast to previous works, great care has been taken to disentangle 
the polarizability from additional effects due to constant electromagnetic fields
on a finite lattice, resulting in $\alpha^n_E=-2.0(9)\cdot10^{-4}$ fm$^3$ 
for the electric polarizability of the neutron at $m_\pi\simeq 760$ MeV.
The substantial difference to the experimental value $\alpha^n_E=11.6(1.5)\cdot10^{-4}$ fm$^3$
may be explained in the framework of chiral perturbation theory, which predicts a strong
pion mass dependence of the form $1/m_\pi$ at low $m_\pi$.
For comparison, we refer to a previous investigation in the 
quenched approximation in Ref.~\cite{Christensen:2004ca}.

\section{Moments of parton distribution functions}

\subsection{Quark momentum fractions in the nucleon}
\label{sec:nuclmomfrac}
The momentum fractions carried by the quarks and gluons in the nucleon, $\langle x\rangle_{q,g}$,
represent fundamental hadron structure observables. They are directly related to
the moments of structure functions through the operator product expansion (OPE).
Although separately dependent on the
scale, $\mu$, and the renormalization scheme, they must add up to one according to the 
fundamental momentum sum rule, $\sum_{q} \langle x\rangle_{q}+\langle x\rangle_{g}=1$, 
independent of the scheme and scale. 
The quark momentum fractions in particular are very well known most notably 
from deep inelastic scattering (DIS) experiments in combination with
phenomenology, and have been studied already for a long time in lattice QCD. 

\begin{figure}[t]
     \begin{minipage}{0.38\textwidth}
      \centering
          \includegraphics[width=0.99\textwidth,angle=0]{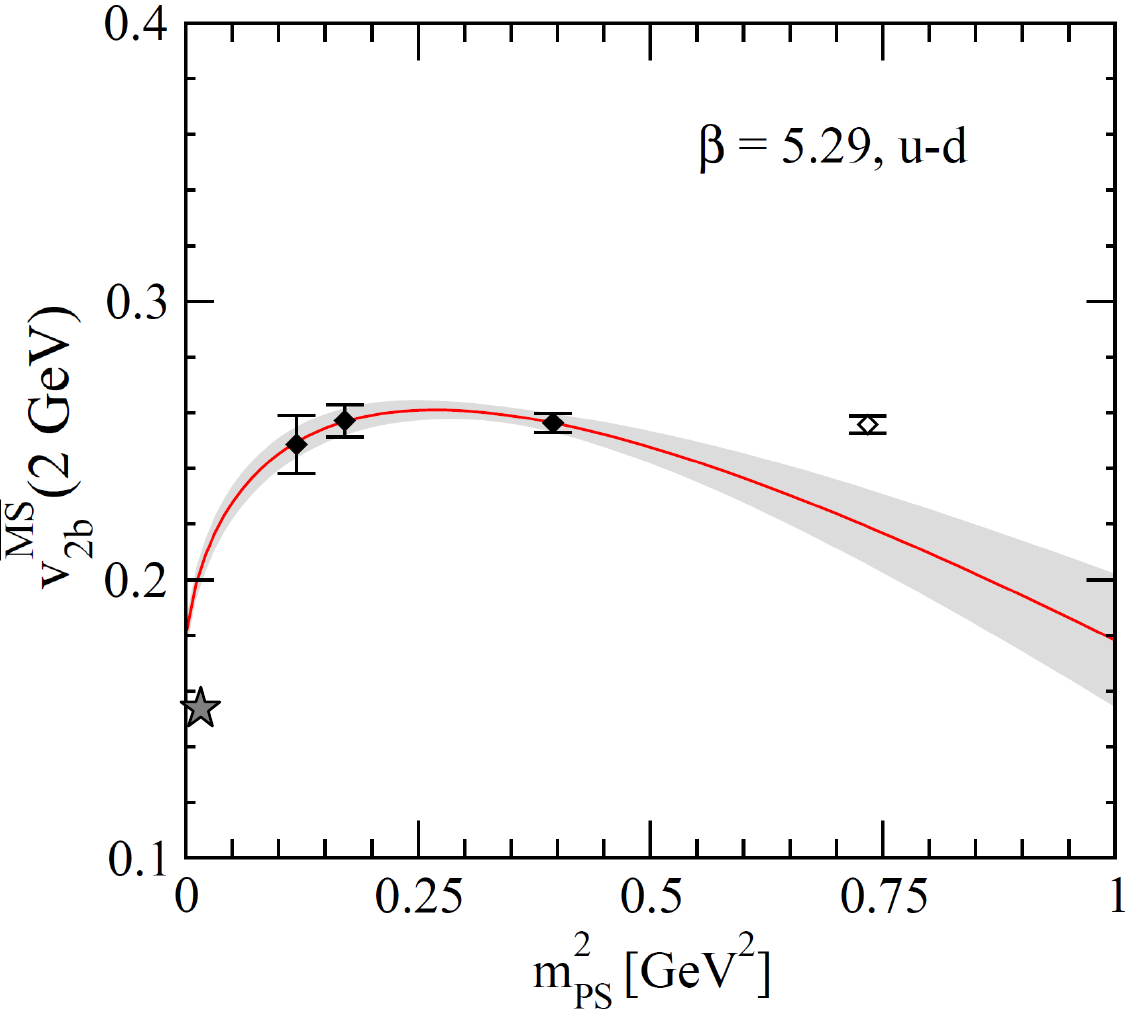}
  \caption{Quark momentum fraction $\langle x\rangle_{u-d}$ in the 
  nucleon from QCDSF/UKQCD\cite{Pleiter:PoSLat2007}.\newline}\label{xQCDSF}
     \end{minipage}
     \hspace{0.5cm}
     \begin{minipage}{0.57\textwidth}
      \centering
          \vspace*{-0.7cm}
          \includegraphics[width=0.9\textwidth,clip=true,angle=0]{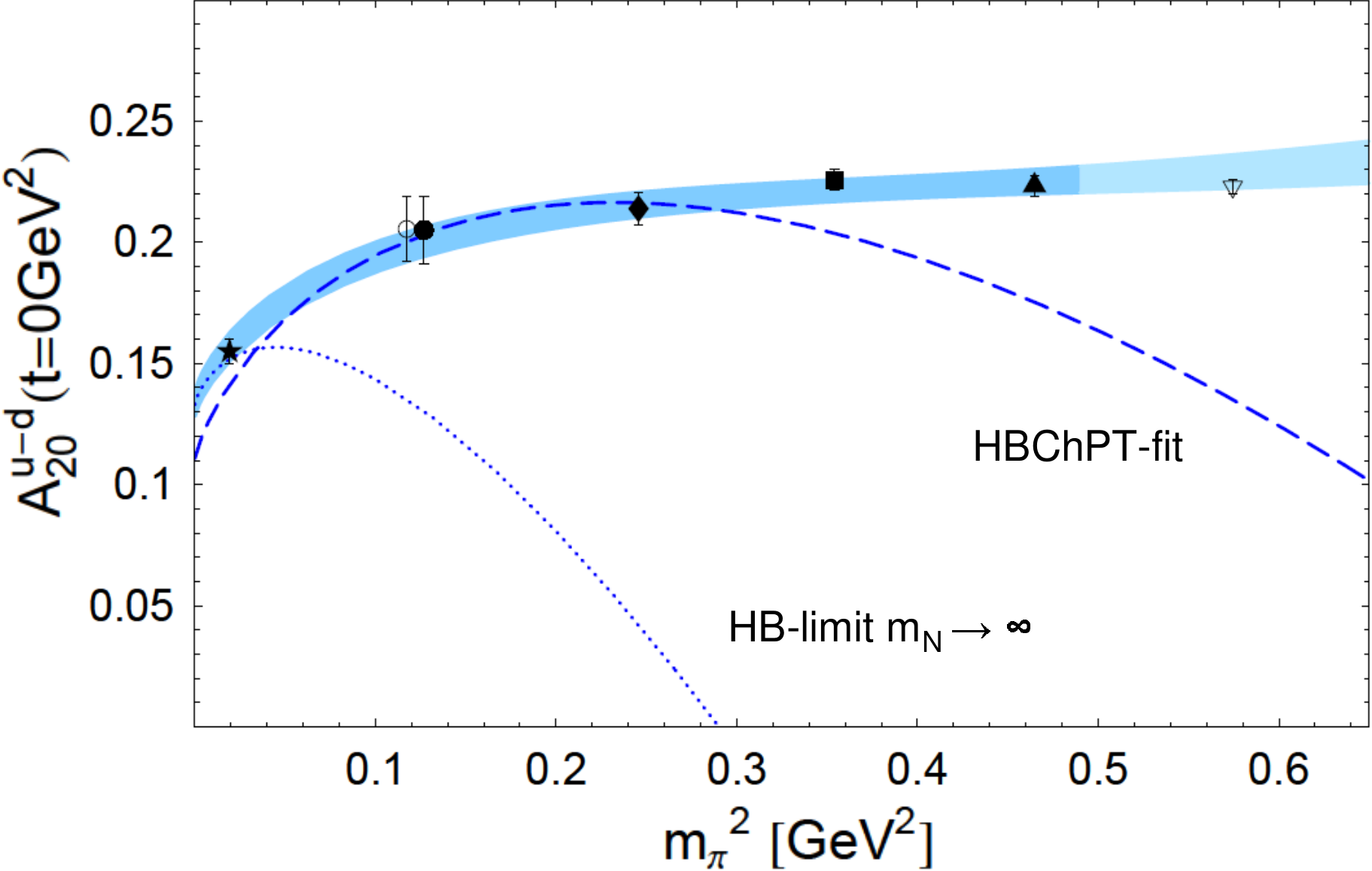}
  \caption{Quark momentum fraction $\langle x\rangle_{u-d}$ in the 
  nucleon from LHPC \cite{Hagler:2007xi,Renner:PoSLat2007}.}
  \label{xLHPC}
     \end{minipage}
     \vspace*{-0.4cm}
   \end{figure}

As discussed in the introduction, leading twist continuum operators, Eq.~(\ref{localOps}),
are used to define moments
of (generalized) PDFs. On a discrete space-time lattice, the corresponding traceless and (anti-) symmetrized 
operators are classified according to irreducible representations of the hypercubic group $H(4)$. 
Typical lattice operators related to the quark momentum fractions are 
$\mathcal{O}_{\text{v}2a}=\bar q \gamma_{\{1}\Dlr_{4\}}q$ and 
$\mathcal{O}_{\text{v}2b}=\bar q (\gamma_{\{4}\Dlr_{4\}}-1/3\sum_{i=1}^3 \gamma_{\{i}\Dlr_{i\}} )q$,
belonging to the representations $\tau^6_3$ and $\tau^3_1$, respectively.
The corresponding lattice quark momentum fractions are often denoted by $\text{v}_{2a}$ and $\text{v}_{2b}$.
In the continuum limit, both operators should of course yield the same result.
Figure~\ref{xQCDSF} shows preliminary results from QCDSF/UKQCD \cite{Pleiter:PoSLat2007} for the isovector 
momentum fraction $\text{v}^{u-d}_{2b}$ versus the pion mass squared at a fixed coupling $\beta=5.29$.
The operator has been non-perturbatively renormalized following the Rome-Southampton method
and transformed to the $\overline{\text{MS}}$ scheme at a scale of $4$ GeV$^2$. Since the dependence on
the lattice spacing appears to be small \cite{Pleiter:PoSLat2007}, we may directly compare with 
results from phenomenology, e.g. $\langle x\rangle^{\text{CTEQ6}}_{u-d}\simeq0.155$, indicated by
the star. At the lowest pion mass of $\sim340$ MeV, the lattice data point is $\approx40\%$
above the CTEQ value. In order to see wether this gap can be bridged, recent results from
covariant baryon ChPT (CBChPT) \cite{Dorati:2007bk,Dorati:PoSLat2007} 
have been used to fit the lattice data. One advantage of the covariant formalism
is that all powers of the ratio $(m_\pi/m_N)$ are resummed, in contrast
to the more widely used non-relativistic heavy baryon approach, which is based on a
\emph{simultaneous} expansion in $m_\pi/\Lambda_\chi$ and $m_\pi/m_N$, 
where $\Lambda_\chi\!\sim\!1$ GeV is the chiral symmetry breaking scale.
The result of a 2-parameter CBChPT-fit to the lattice data is shown as shaded error band in Fig.~\ref{xQCDSF}.
Although a rather strong downwards bending is visible,
the fit misses the phenomenological value at the physical point by $\approx20\%$.

Results for the isovector momentum fraction by LHPC \cite{Hagler:2007xi,Renner:PoSLat2007}, obtained
in the framework of an extensive study of moments of GPDs, are presented in Fig.~\ref{xLHPC}. 
Here it is useful to note that the momentum fraction is equal to the $(n=2)$-moment 
of the vector GPD $H$ in the forward limit, 
$\int dx x H(x,0,t=0)=A_{20}(t=0)=\langle x\rangle$. The calculation is based on a combined analysis of the perturbatively renormalized (with NP-improvement, see \cite{Hagler:2007xi}) 
lattice operators in the representations $\tau^6_3$ and $\tau^3_1$.  The shaded error band is
 the result of a {\it global} CBChPT-fit to the lattice data for $m_\pi^2<0.47$ GeV$^2$ and momentum transfers
$|t|\le0.47$ GeV$^2$. A very good agreement with the phenomenological value
is found at the physical pion mass. This may be attributed to the overall lower normalization
of the lattice data points, in particular compared to the results in Fig.~\ref{xQCDSF}, 
as well as the inclusion of all powers $(m_\pi/m_N)^n$ in the CBChPT approach, 
compared to the heavy baryon formalism, as can be seen from the dotted and dashed lines in Fig.~\ref{xLHPC}. 
Even taking into account that the calculations presented in \cite{Pleiter:PoSLat2007} (Fig.~\ref{xQCDSF}) 
and \cite{Hagler:2007xi} (Fig.~\ref{xLHPC})
are based on different lattice actions and methods, the difference in 
the normalization of $\langle x\rangle_{u-d}$ can be regarded as substantial
and certainly demands a closer examination, including possible finite size effects \cite{Detmold:2005pt}.

New results on isovector momentum fractions in the nucleon 
based on dynamical DW fermions have been presented by RBC-UKQCD \cite{Yamazaki:PoSLat2007}.

In addition to isovector quantities, which provide essential information on hadron structure,
calculations in the isosinglet and the strange-quark sector are required for the analysis of a large number
of important observables and e.g. the momentum and spin sum rules.
However, the numerically expensive contributions from disconnected diagrams are mostly 
neglected so far, and such calculations are therefore subject to systematic uncertainties of largely unknown size.
Preliminary results on the disconnected contributions to the quark momentum fraction in the nucleon 
have been obtained in a recent effort by Deka and Liu \cite{Deka:xyz2007}, using the standard Wilson
gauge and fermion action in the quenched approximation for pion masses down to $480$ MeV.
Their calculation of the all-to-all propagators is based on the $Z_2$ noise estimator and 
an unbiased subtraction to reduce the variance using the hopping parameter expansion.
Employing perturbatively renormalized operators of the off-diagonal type, e.g. $\mathcal{O}_{\text{v}2a}$,
they find in the $\overline{\text{MS}}$ scheme at a scale of $\sim3$ GeV$^2$  
a strange quark contribution to the momentum fraction of $\langle x\rangle_{s}=0.046(17)$,
and $\langle x\rangle^{\text{disc}}_{u,d}=0.055(18)$ for the contribution from disconnected insertions.
These results have been linearly extrapolated to the chiral limit. This shows that
disconnected diagrams may give a substantial, non-negligible contribution to the
total quark momentum fraction as large as $\sim10\%$
of the value for the connected part in the quenched approximation, 
$\langle x\rangle^{\text{con}}_{u+d}\approx0.65$.
Studies along similar lines have been been presented by Babich \cite{Babich:PoSLat2007},
using stochastic noise estimators to compute the strange quark contribution to the axial-vector and scalar form factor of the nucleon, 
and Collins \cite{Collins:PoSLat2007}, 
discussing the use of stochastic methods combined with unbiased subtraction and a truncated
inversion of the stochastic propagator in the calculation of disconnected contributions to hadronic structure.

A possible alternative to the calculation of moments of PDFs using local operators is the direct study of the OPE
on the lattice. Interesting preliminary results in this direction based on overlap fermions 
have been presented by QCDSF/UKQCD \cite{Rakow:PoSLat2007}.

\subsection{Parton momentum fractions in the pion}
\begin{figure}[t]
     \begin{minipage}{0.4\textwidth}
      \centering
          \includegraphics[width=0.9\textwidth,clip=true,angle=0]{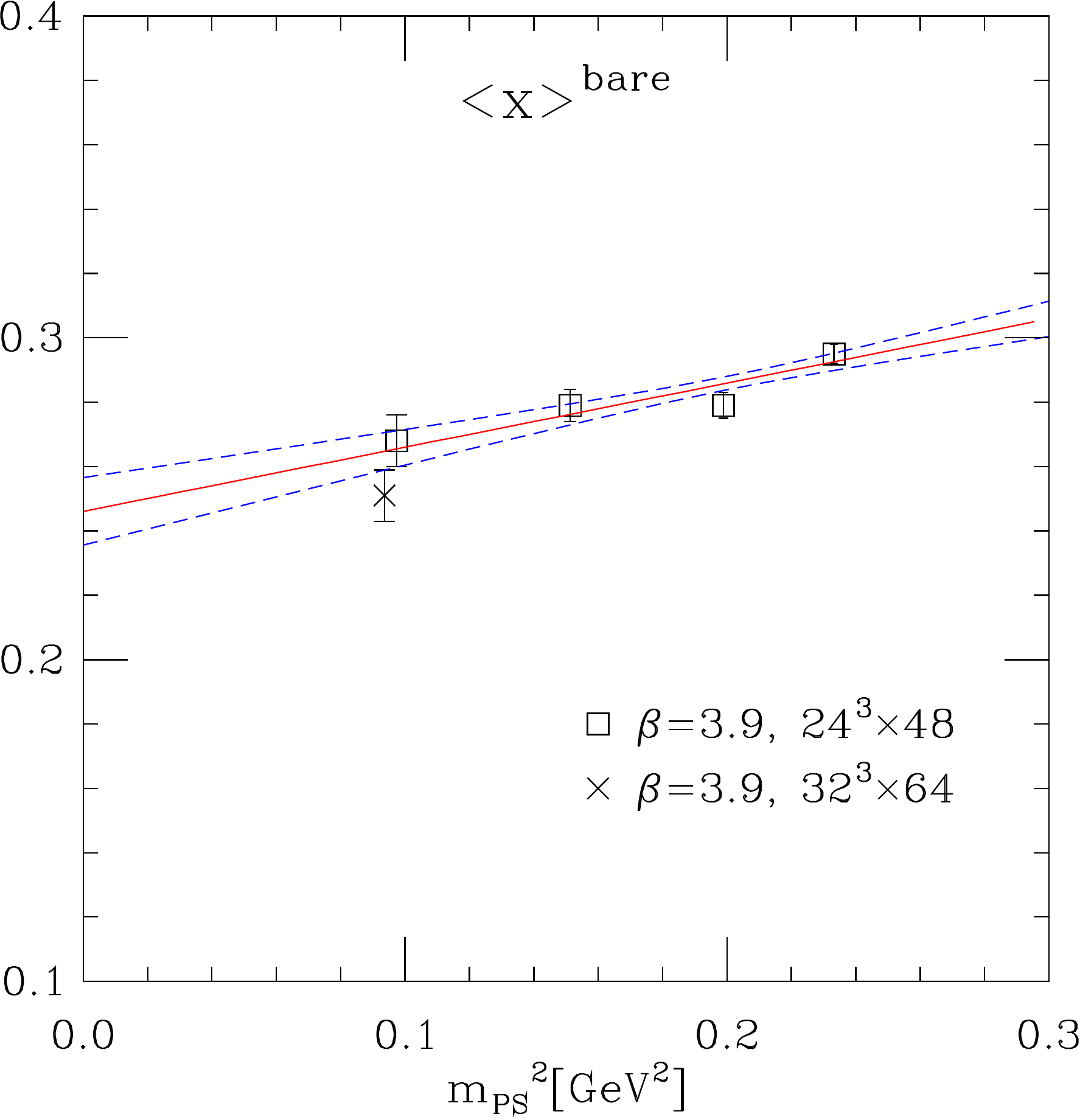}
      \vspace*{-0.3cm}
  \caption{Bare quark momentum fraction in the pion $\langle x\rangle^\pi$ from 
   ETMC \cite{ZLiu:PoSLat2007}.\newline}\label{xPionETMC}
     \end{minipage}
     \hspace{0.5cm}
     \begin{minipage}{0.55\textwidth}
      \centering
      \vspace*{-0.4cm}
          \includegraphics[angle=-90,width=0.94\textwidth,clip=true]{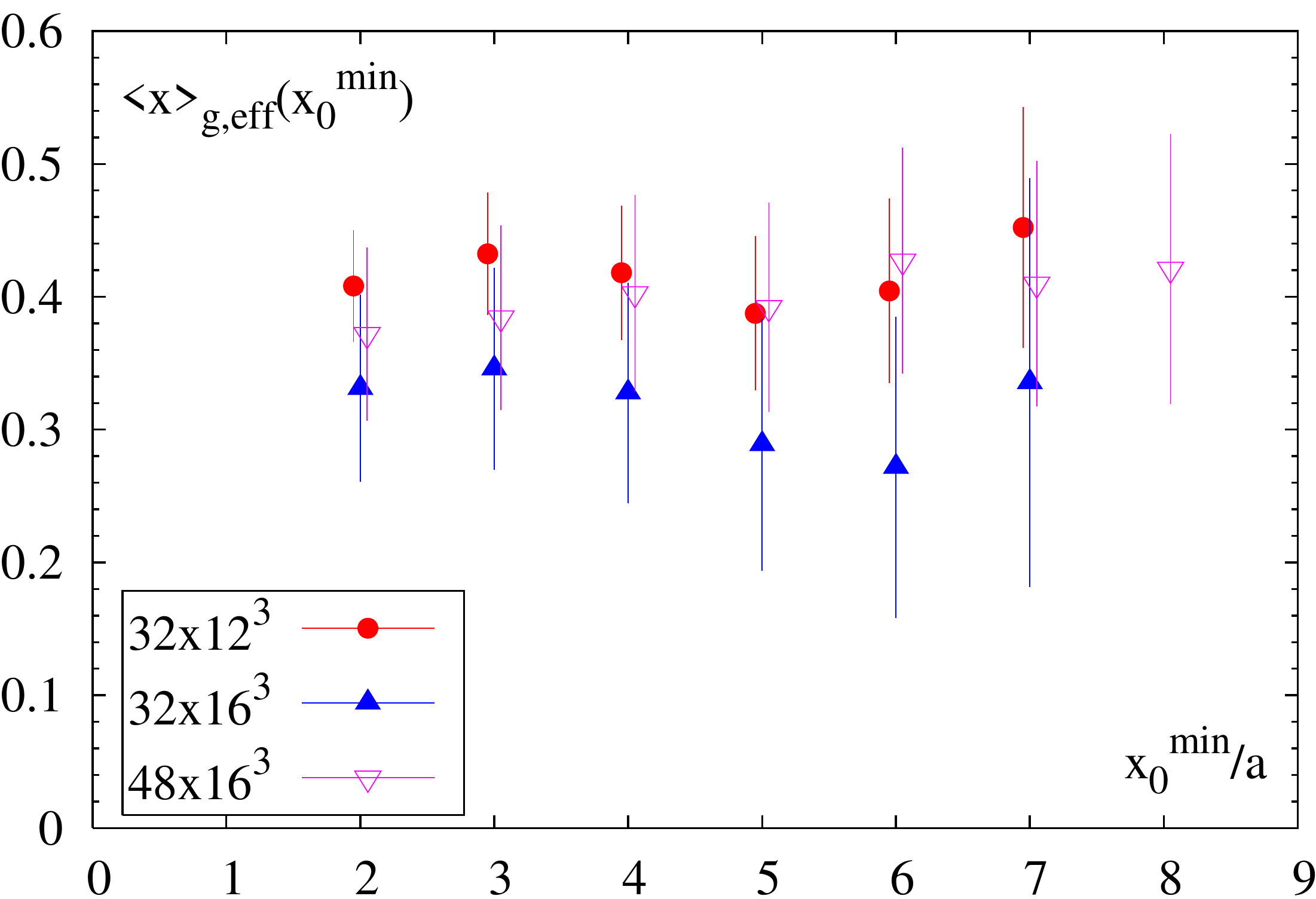}
      \vspace*{0.2cm}
  \caption{Gluon momentum fraction $\langle x\rangle^\pi_g$ in a "heavy" 
  pion with $m_\pi\simeq1.06$ GeV \cite{Meyer:PoSLat2007}.}
  \label{xgMIT}
     \end{minipage}
     \vspace*{-0.4cm}
   \end{figure}
%
%

Remarkable work in progress by ETMC \cite{ZLiu:PoSLat2007} on the momentum fraction
of quarks in the pion, i.e. $\langle x\rangle^{\pi^+}_{u}$, is presented in Fig.~\ref{xPionETMC}.
The calculation is based on the $H(4)$-operator $\mathcal{O}_{\text{v}2b}$ defined in the previous section,
and the pion 2- and 3-point functions have been evaluated using stochastic sources instead of point sources.
Using just $\simeq300$ configurations for the measurement, this still leads to a high numerical precision at the lowest pion mass of $\sim300$ MeV (cross and leftmost square in Fig.~\ref{xPionETMC}), where the statistical error for the unrenormalized quark momentum fraction in the pion is $\mathcal{O}(3\%)$. 
These results, as well as calculations by QCDSF/UKQCD on the pion structure \cite{Brommel:PoSLat2007},
indicate that the quarks in the pion carry around $50-60\%$ of the total pion momentum. 

This may serve as a reminder of the well-known fact that gluons give substantial, if not major, contributions to
such fundamental properties as the mass and momentum of hadrons. The
calculation of gluonic contributions in lattice QCD to hadron structure observables, 
however, poses a long-standing problem, since short-range quantum fluctuations 
lead to noisy signals. Using a large number of configurations in a quenched Wilson fermion and Wilson glue simulation, 
this problem was tackled in a pioneering exploratory study of the gluon momentum fraction in the nucleon
in Ref.~\cite{Gockeler:1996zg} more than ten years ago. Recently, Meyer and Negele \cite{Meyer:PoSLat2007,Meyer:2007tm}
have studied the gluon momentum fraction in the pion, $\langle x\rangle^{\pi}_{g}$, 
in quenched lattice QCD using the Wilson action. 
In order to reduce the gauge-field fluctuations, they employed HYP smearing to smooth the fields. 
By comparing bare-clover, HYP-plaquette and HYP-clover discretizations of the gluon energy momentum tensor,
they found that the HYP-plaquette version reduces the variance of the entropy density by a factor of $\sim40$
relative to the bare-plaquette discretization, without leading to uncontrollable nonlocality effects.
Their results for the bare gluon momentum fraction are shown in Fig.~\ref{xgMIT} as a function of
the minimum distance $x_0$ in Euclidean time direction between the gluon operator and the pion source and sink.
Taking into account the mixing of the singlet quark and gluon operators in the renormalization procedure,
they obtain $\langle x\rangle^{\pi}_{g}=0.37\pm0.08_{\text{stat}}\pm0.12_{\text{sys}}$ in the
$\overline{\text{MS}}$ scheme at a scale of 4 GeV$^2$ and a pion mass of $m_\pi\simeq0.89$ GeV. 
The estimated systematic error of $\pm0.12$ originates from the fermion normalization constant $Z_f$ which
is so far not known beyond the (trivial) tree level value, $Z_f=1+\mathcal{O}(g^2)$.
Together with corresponding results from \cite{Guagnelli:2004ga} for the connected quark momentum fraction in the
 pion, they find 
$\sum_q \langle x\rangle^{\pi,\text{con}}_{q}+\langle x\rangle^{\pi}_{g}=0.99\pm0.08_{\text{stat}}\pm0.12_{\text{sys}}$, 
very close to one, but still allowing for disconnected quark contributions as large as $10\%$ within errors.

\subsection{Moments of meson distribution amplitudes}
\begin{figure}[t]
\bc
\includegraphics[width=9.cm,angle=0,clip=true]{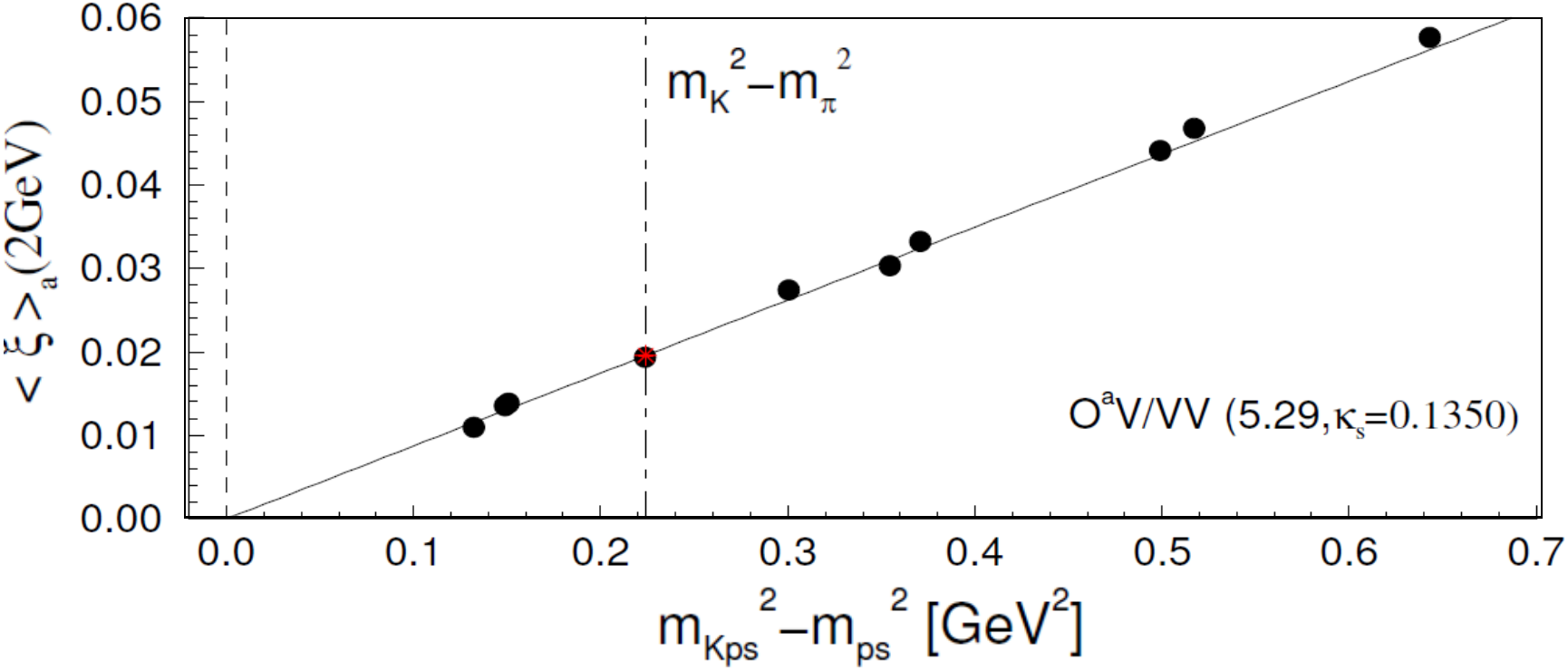}
\caption{\label{xiKstar} Moment of the $K^*$ DA $\langle \xi\rangle_{K^*}$ from QCDSF/UKQCD \cite{Horsley:PoSLat2007}.}
\ec 
\vspace*{-0.5cm}
\end{figure}
Important information on hadron structure for small transverse parton
separations is provided by hadronic distribution amplitudes (DAs).
Meson DAs are denoted by $\phi(\xi)$, where $\xi$ is directly related to the quark and anti-quark
longitudinal momentum fractions through $x=1/2(1+\xi)$ and $1-x=1/2(1-\xi)$, respectively.
Following calculations of the lowest two moments of pseudo-scalar meson DAs \cite{Braun:2006dg},
QCDSF/UKQCD now presented first results for moments of vector meson DAs, specifically the $K^*$,
in lattice QCD \cite{Horsley:PoSLat2007}.
The DA $\phi_{K^*}(\xi)$ is an essential ingredient in the description of rare $B$ decays like
$B\to  K^*l^+l^-$ in the framework of QCD factorization. Since these decays are induced by 
flavor-changing neutral current transitions only occurring through loops in
the standard model, they are an ideal tool to probe new physics. 
Moments of meson DAs, $\langle \xi^{n-1}\rangle=\int d\xi \xi^{n-1}\phi(\xi)$,
parametrize meson-to-vacuum matrix elements of 
the quark operators $\mathcal{O}^{\mu\mu_1\ldots}$ in Eq.~(\ref{localOps}).
For the calculation of the ($n\!=\!2$)-moment $\langle \xi\rangle_{K^*}$, 
the (NP-renormalized) operators  $\mathcal{O}_{\text{v}2a}$ and $\mathcal{O}_{\text{v}2b}$,
discussed in section \ref{sec:nuclmomfrac}, have been used. Figure \ref{xiKstar} shows
the result for $\langle \xi\rangle_{a, K^*}$ at a fixed sea quark mass and coupling versus
the difference $m_{K}^2-m_\pi^2$ in the squared pseudo-scalar lattice masses. The vertical
dashed-dotted line represents the physical value for $m_{K}^2-m_\pi^2$. An interpolation of
$\langle \xi\rangle_{a, K^*}$ to this point, followed by a linear chiral extrapolation in $m_\pi^2$,
gives $\langle \xi\rangle_{a, K^*}\!\approx\!0.036(3)$, in the $\overline{\text{MS}}$ scheme at a scale of 4 GeV$^2$.
This first-time study proves the feasibility of future high statistics dynamical lattice QCD calculations 
of moments of vector meson DAs.

Promising results on moments of pion and kaon DAs in a dynamical DW lattice calculation with pion masses
as low as $300$ MeV have been presented by Sachrajda \cite{Sachrajda:PoSLat2007} for RBC-UKQCD.
Using perturbatively renormalized operators, the preliminary results in the $\overline{\text{MS}}$ scheme 
for $\mu= 2$ GeV are
$\langle \xi^2\rangle_{\pi}\approx0.28(3)$ for the pion, and
$\langle \xi\rangle_{K}\approx0.029(2)$, $\langle \xi^2\rangle_{K}\approx0.27(2)$ for the kaon,
showing that "lattice calculations can be performed with an excellent precision" \cite{Sachrajda:PoSLat2007}.

\section{Moments of generalized parton distributions}
\hspace{-3mm}A number of recent lattice results on moments of GPDs have been presented
by QCDSF/UKQCD for the nucleon \cite{Ohtani:PoSLat2007} and the pion \cite{Brommel:PoSLat2007} 
and by LHPC for the nucleon \cite{Renner:PoSLat2007}.
Below, we discuss only a small selection lattice of these results, related to the nucleon and pion spin structure.

\subsection{Decomposition of the nucleon spin\label{SecSpin}}
\begin{figure}[t]
      \begin{minipage}{0.48\textwidth}
      \centering
      \vspace*{-0.2cm}
          \includegraphics[angle=-90,width=0.9\textwidth,clip=true]{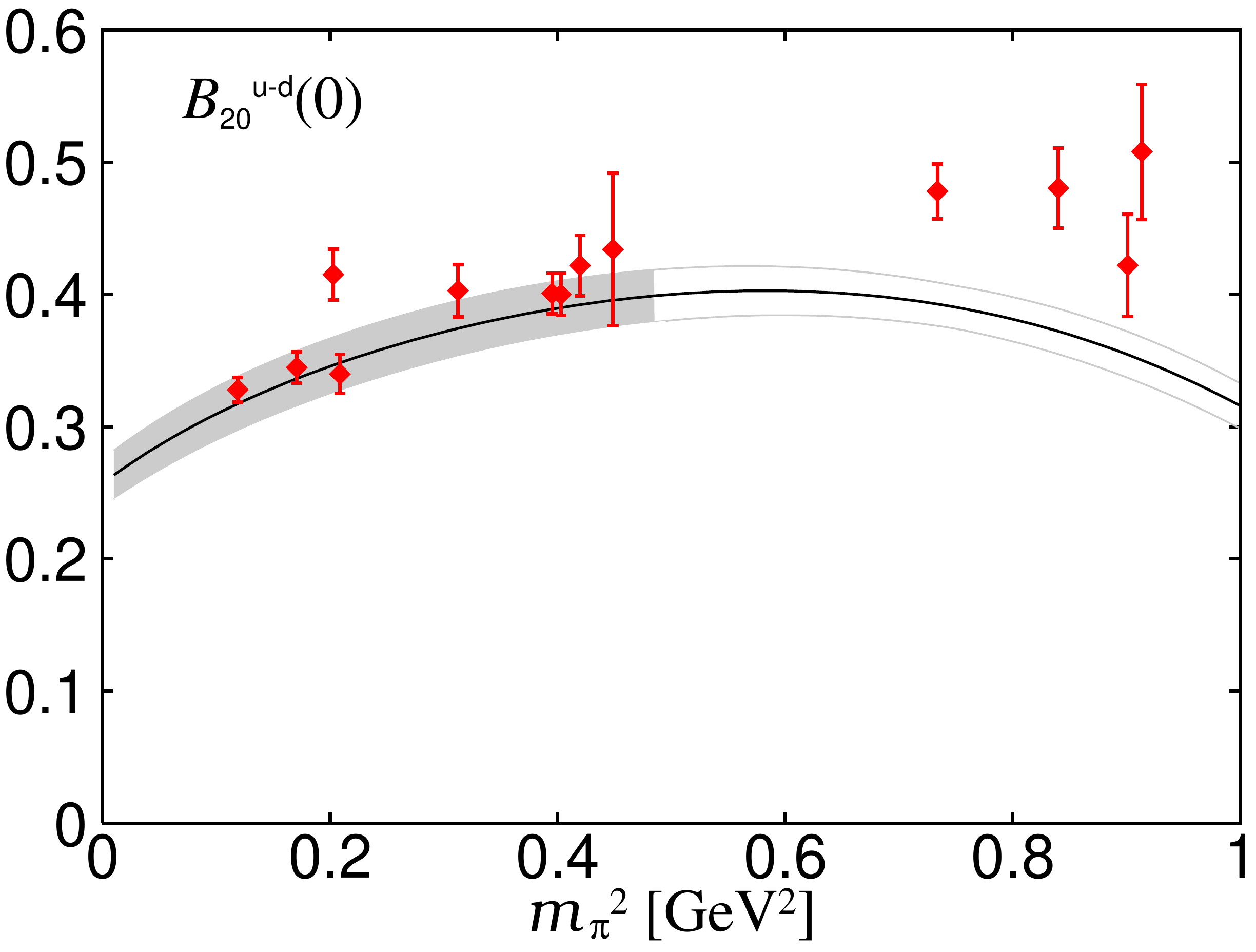}
  \caption{The GFF $B^{u-d}_{20}(t\!=\!0)$ from QCDSF/UKQCD \cite{Ohtani:PoSLat2007}.}
  \label{B20QCDSF}
     \end{minipage}
     \hspace{0.4cm}
     \begin{minipage}{0.48\textwidth}
      \centering
           \includegraphics[angle=-90,width=0.9\textwidth,clip=true]{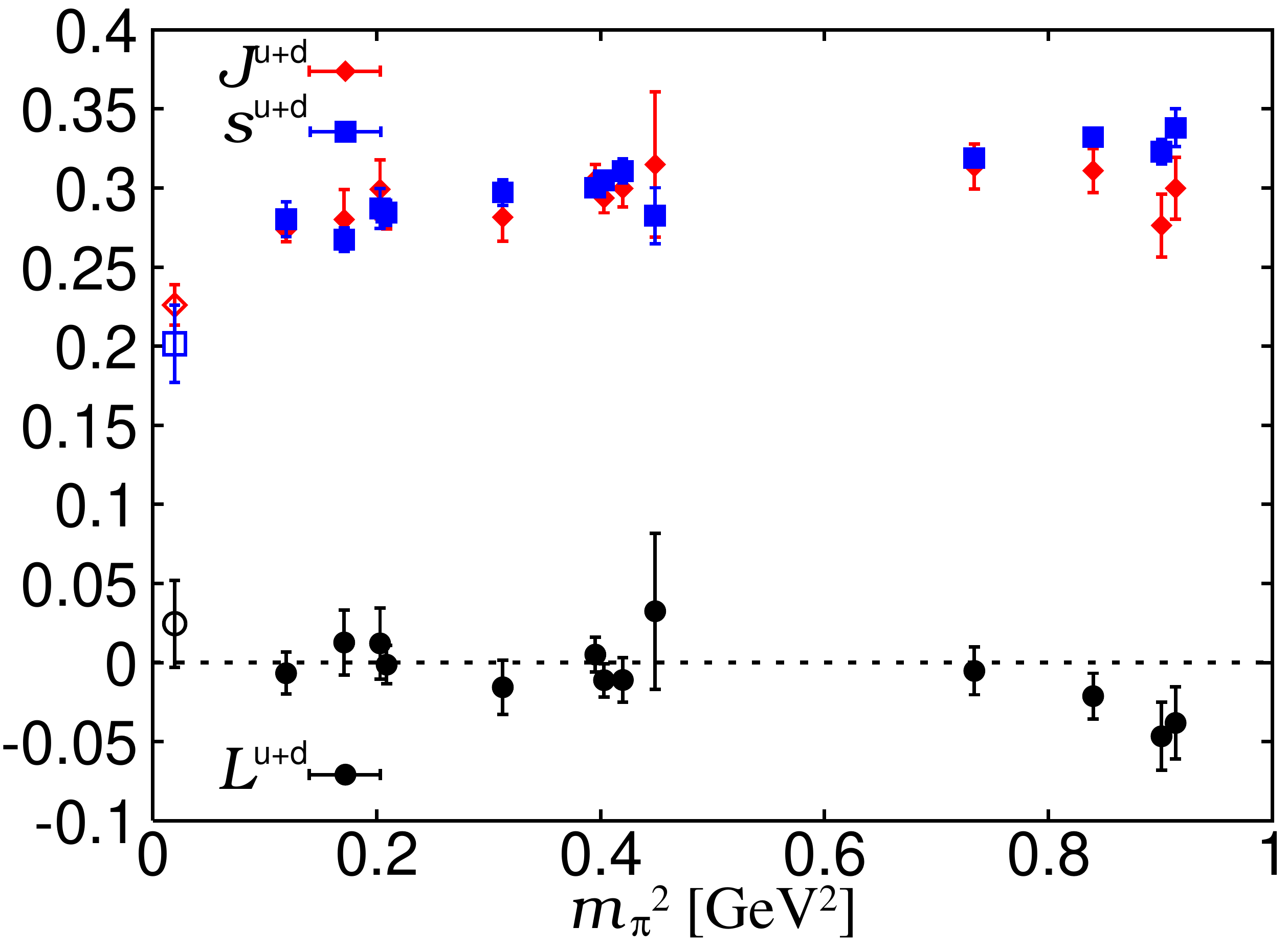}
  \caption{Decomposition of the nucleon spin from QCDSF/UKQCD \cite{Ohtani:PoSLat2007}.\newline}\label{OAMQCDSF}
     \end{minipage}
     \vspace*{-0.5cm}
   \end{figure}  

Of particular interest in the investigation of the structure of the nucleon
is the decomposition of the nucleon spin $1/2$ in terms of quark and gluon
spin and orbital angular momentum (OAM) contributions. According to the nucleon spin sum rule , we have
\begin{equation}
\frac{1}{2}=\frac{1}{2}\Delta\Sigma+\Delta G + L_q + L_g\,,
\end{equation}
where $\Delta\Sigma$ and $\Delta G$ are the standard gauge-invariant quark and
gluon spin fractions, while the OAM are defined by $L_{q}=J_q-\Delta\Sigma/2$
and $L_{q}=J_g-\Delta G$. The gauge-invariant total angular momenta 
$J_{q,g}$ are related to the forward values of the $(n=2)$-moments of 
the GPDs $H$ and $E$ through $2J_{q,g}=\int dxx(H(x,\xi,0)+E(x,\xi,0))=(A^{q,g}_{20}(0)+B^{q,g}_{20}(0))$ \cite{Ji:1996ek}.
It is important to note that the decomposition of the nucleon spin is in general 
scale and scheme dependent.
Since the $A^{q,g}_{20}(t=0)=\langle x\rangle_{q,g}$ are just the well-known quark and gluon
momentum fractions, the genuine contribution from GPDs to the total nucleon spin is
given by the GFFs $B^{q,g}_{20}$. Figure \ref{B20QCDSF} shows preliminary results from
QCDSF/UKQCD for the isovector $B^{u-d}_{20}(t=0)$ as a function of the pion mass squared \cite{Ohtani:PoSLat2007}.
The lattice data points have been chirally extrapolated based on
results from the covariant BChPT calculation presented in \cite{Dorati:2007bk,Dorati:PoSLat2007},
as indicated by the shaded band. A sizeable contribution of $B^{u-d}_{20}(t=0)=0.269(20)$ 
is found at the physical point. Corresponding results in the isosinglet channel, together with
preliminary results for $\Delta\Sigma^{u+d}$ and  $\langle x\rangle_{u+d}$  lead to
the nucleon spin decomposition shown in Fig.~\ref{OAMQCDSF}, 
in the $\overline{\text{MS}}$ scheme at a scale of 4 GeV$^2$.
It is remarkable that the $u+d$-quark OAM 
\begin{wrapfigure}{r}{0.5\textwidth}
  \begin{center}
  \hspace*{-0.2cm}
  \vspace*{-0.5cm}
      \includegraphics[width=0.51\textwidth]{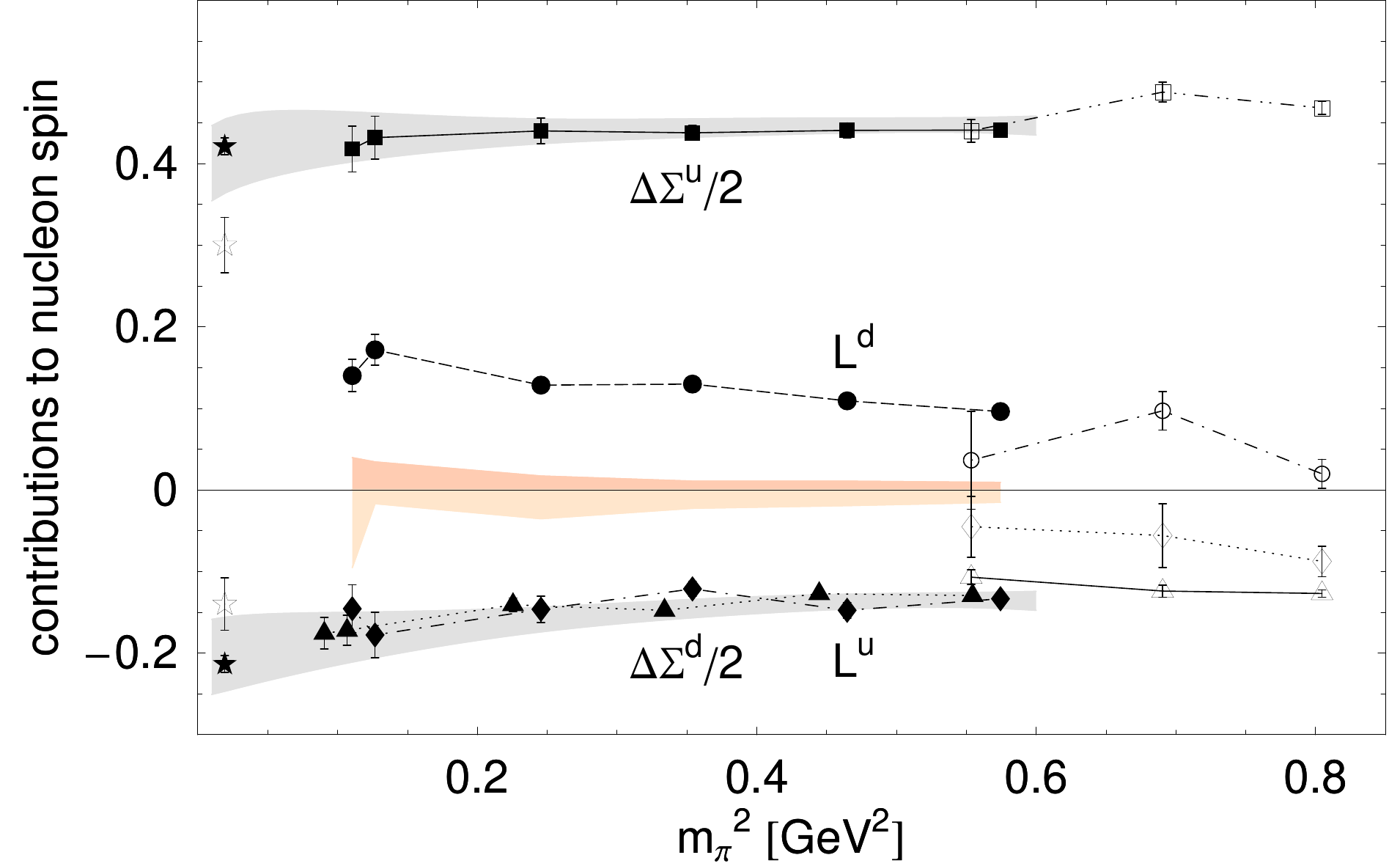}
  \end{center}
  \vspace*{-0.2cm}
 \caption{Decomposition of the nucleon spin from LHPC \cite{Hagler:2007xi}. Filled
 stars represent recent HERMES results for $\Delta\Sigma^{u,d}/2$ \cite{Airapetian:2007mh}.}
  \label{OAMLHPC}
\end{wrapfigure}
\noindent 
contribution is compatible with zero  over the full range of accessible pion masses. 
A further decomposition in terms of up and down quark spin and OAM contributions to the
nucleon spin by LHPC \cite{Hagler:2007xi,Renner:PoSLat2007} is presented in Fig.~\ref{OAMLHPC}.
Interestingly, the separate $u$ and $d$ quark OAM contributions are quite sizeable but opposite in sign,
$L_d\simeq-L_u\approx30\%$ of $1/2$, and only cancel in sum, $L_{u+d}\approx0$, for all pion masses.
Similarly, the down quark spin and OAM contributions in Fig.~\ref{OAMLHPC} are of same size
but opposite in sign, so that the total down quark angular momentum is approximately zero, 
$J_d=\Delta\Sigma_d/2+L_d\approx0$. The chiral extrapolations in \cite{Hagler:2007xi} indicate 
that this also holds at the physical point, so that the total quark contribution to the spin
of the nucleon is to nearly 100\% coming from the up quarks. All these results should be taken
with due caution, since disconnected contributions have not been included in the calculations
presented in  Fig.~\ref{OAMQCDSF} and Fig.~\ref{OAMLHPC}. 
A comparison with recent combined experimental results on $J_{u,d}$ from JLab and
HERMES can be found in \cite{collaboration:2007vj}.

\subsection{Spin structure of the pion}
\begin{figure}[t]
      \begin{minipage}{0.38\textwidth}
      \centering
      \vspace*{-0.2cm}
          \includegraphics[width=0.65\textwidth,clip=true,angle=0]{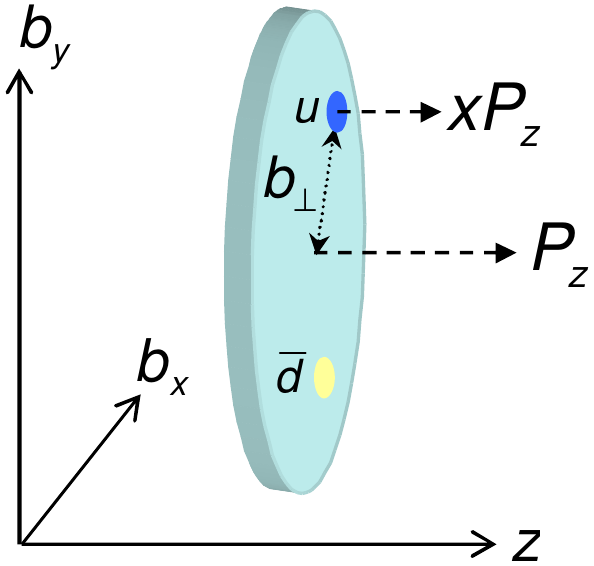}
  \caption{Density interpretation of GPDs in impact parameter space.}
  \label{ImpactSpace}
     \end{minipage}
     \hspace{0.4cm}
     \begin{minipage}{0.58\textwidth}
      \centering
       \includegraphics[width=0.98\textwidth]{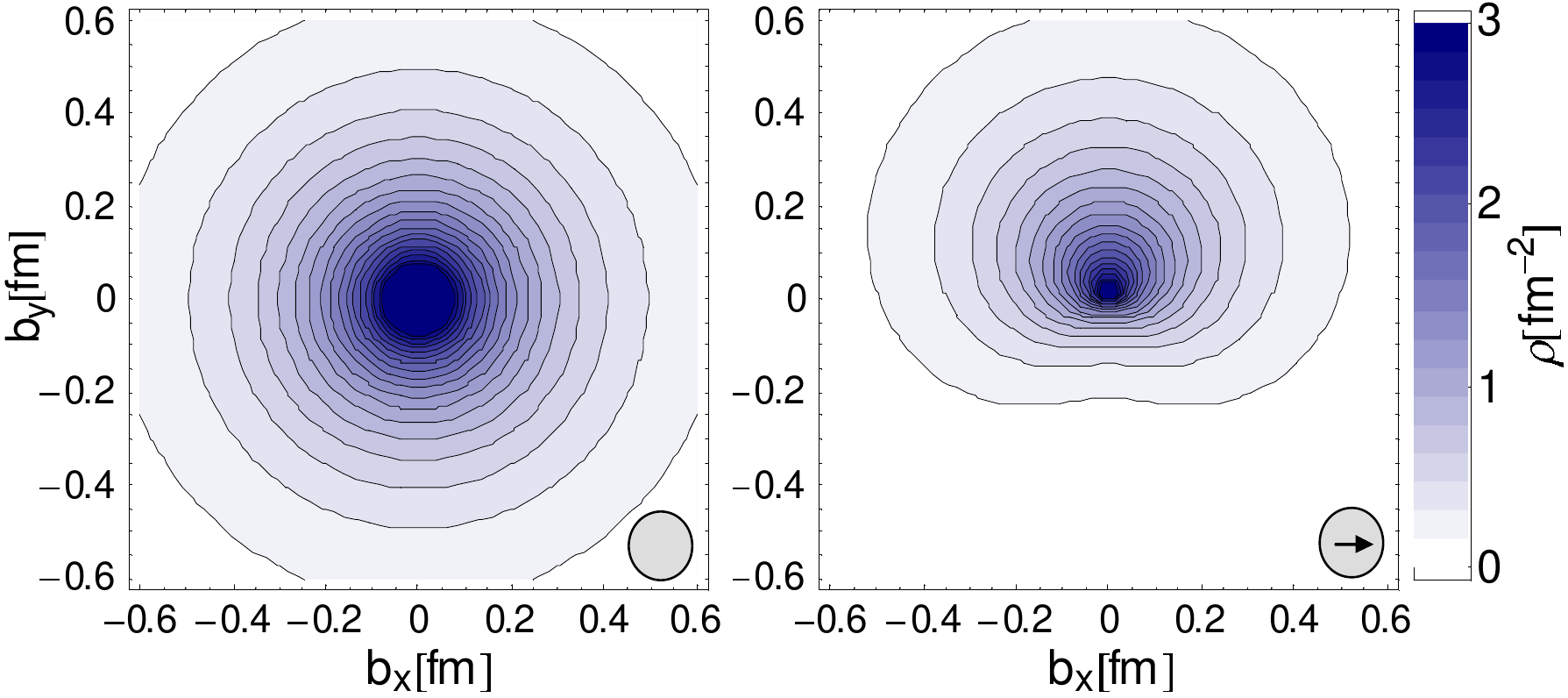}
    \caption{\label{PionDensityQCDSF}Densities of up-quarks in the $\pi^+$ from QCDSF/UKQCD
    \cite{Brommel:PoSLat2007,Brommel:2007xd}. The arrow indicates the orientation of 
    the transverse quark spin $s_\perp$.}
     \end{minipage}
   \end{figure}

A longstanding, important question is how quarks are spatially distributed inside hadrons.
In \cite{Diehl:2005jf}, it has been shown that vector and tensor GPDs in particular allow for a proper definition
of transverse coordinate (impact parameter) space densities of transversely polarized quarks in the nucleon.
Impact parameter densities, which can be determined from Fourier transforms
of GPDs with respect to the transverse momentum transfer, 
$H(x,b_\perp^2)\!=\!(2\pi)^{-2}\int d^2\Delta_\perp\exp(\text{{\ttfamily -}}i\Delta_\perp b_\perp)H(x,0,\text{{\ttfamily -}}\Delta_\perp^2)$,
have been introduced by Burkardt \cite{Burkardt:2000za} and are illustrated in Fig.~\ref{ImpactSpace}.
In analogy to the nucleon case \cite{Gockeler:2006zu}, the density $\rho(x,b_\perp,s_\perp)$ 
of quarks with longitudinal momentum fraction $x$, transverse spin $s_\perp$ 
at transverse distance $b_\perp$ of the center of momentum in the pion 
 can be directly obtained from the pion vector and tensor GPDs, 
 $H^\pi(x,b_\perp^2)$ and $E^\pi_T(x,b_\perp^2)$, respectively.
The lowest two moments of these pion GPDs have been calculated recently by QCDSF/UKQCD 
\cite{Brommel:PoSLat2007,Brommel:2007xd}, and the
corresponding results for the lowest $x$-moment of the density, $\rho^{n=1}(b_\perp,s_\perp)$,
are shown in Fig.~\ref{PionDensityQCDSF} for up quarks in a $\pi^+$. 
Compared to the unpolarized case on the left, the
density of quarks with transverse spin in $x$-direction on the right in Fig.~\ref{PionDensityQCDSF}
is strongly distorted in $b_y$-direction. This result is very similar to what has been obtained
for quarks in the nucleon \cite{Gockeler:2006zu} and proves the presence of strong correlations
of the form $s_\perp\times b_\perp$ between transverse spin and coordinate degrees of freedom of quarks in the pion.
Following arguments by Burkardt \cite{Burkardt:2005hp} and others \cite{Meissner:2007rx}, 
this distortion also indicates that the so-called Boer-Mulders function $h^{\pi,\perp}_1(x,k_\perp)$ of the pion, 
describing correlations of the intrinsic quark transverse momentum, $k_\perp$, and the transverse quark spin, 
is large and negative. This may be important for future studies of azimuthal asymmetries
in polarized and unpolarized $\pi p$ Drell-Yan production at COMPASS/CERN.

Preliminary results of a first direct study of transverse momentum dependent PDFs 
on the lattice based on non-local operators have been presented by Musch \cite{Musch:PoSLat2007}.

\section{Summary}
Recent lattice hadron structure calculations have provided substantial new insights
into the shape of hadrons, their momentum and spin structure in terms 
of quarks and gluons and the spatial distribution of quarks in 
the nucleon and the pion. 
As a consequence, an increasing number of lattice results in this field have a direct impact 
on current and future experimental measurements and related phenomenology.
Remarkable progress has been made in calculations of the pion structure based on 
dynamical chiral fermions, using all-to-all propagators as well as partially twisted boundary conditions. 
It will be interesting to see if this can be repeated in the near future for some 
of the more involved nucleon structure observables.
Despite these successes, several questions remain to be answered, regarding, e.g., 
discrepancies between different computations of the isovector quark momentum 
fraction in the nucleon, the consistently low lattice results for mean square 
charge radii, and the importance of disconnected contributions in dynamical simulations.
Finally, as we enter the regime of precision calculations at pion masses $\simeq 300$ MeV
and below, chiral extrapolations based on ChEFT of QCD become increasingly
relevant and will help to significantly reduce systematic uncertainties of the lattice computations.

\begin{acknowledgments}
It is a pleasure to thank C.~Alexandrou, D.~Br\"ommel, S.~Collins, M.~Engelhardt, M.~G\"ockeler,
R.~Horsley, A.~J\"uttner, T.~Kaneko, K.-F.~Liu, Z.~Liu, B.~Musch, J.W.~Negele, M.~Ohtani, D.~Pleiter, 
M.~Procura, D.B.~Renner, A.~Sch\"afer, G.~Schierholz, W.~Schroers, S.~Simula, 
T.~Yamazaki and J.~Zanotti, 
for helpful correspondence, providing exciting results and figures. 
I gratefully acknowledge the support by the Emmy-Noether program of the DFG. 
\end{acknowledgments}

\end{document}